\documentclass[apj]{emulateapj}

\usepackage{times}
\usepackage{amssymb}
\usepackage{amsmath}
\usepackage{pifont}
\usepackage{graphics}
\usepackage[svgnames]{xcolor}
\usepackage{epsfig}
\usepackage{threeparttable}

\usepackage[normalem]{ulem}
\usepackage{color}

\newcommand\redsout{\bgroup\markoverwith{\textcolor{red}{\rule[0.5ex]{2pt}{1.0pt}}}\ULon}

\newcommand{\MJ}{M\textsubscript{J}}          
\newcommand{\RP}{R\textsubscript{P}}          
\newcommand{\Mstar}{M\textsubscript{$\star$}} 
\newcommand{\Rstar}{R\textsubscript{$\star$}} 

\newcommand{\bjdtdb}{\ensuremath{\rm {BJD_{TDB}}}}
\newcommand{\feh}{\ensuremath{\left[{\rm Fe}/{\rm H}\right]}}
\newcommand{\mh}{\ensuremath{\left[{\rm m}/{\rm H}\right]}}
\newcommand{\teff}{\ensuremath{T_{\rm eff}\,}}

\newcommand{\msun}{\ensuremath{\,M_\Sun}}
\newcommand{\rsun}{\ensuremath{\,R_\Sun}}
\newcommand{\lsun}{\ensuremath{\,L_\Sun}}

\newcommand{\av}{\ensuremath{A_V}}
\newcommand{\dstar}{\ensuremath{d_\star}}
\newcommand{\mj}{\ensuremath{\,M_{\rm J}}}
\newcommand{\rj}{\ensuremath{\,R_{\rm J}}}
\newcommand{\fave}{\langle F \rangle}
\newcommand{\fluxcgs}{10$^9$ erg s$^{-1}$ cm$^{-2}$}

\newcommand{\kms}{\,km\,s$^{-1}$}

\newcommand{\loggstar}{\ensuremath{\log{g_\star}}}
\newcommand{\teq}{\ensuremath{T_{\rm eq}}}
\newcommand{\vsini}{\ensuremath{v\sin{I_*}}}

\newcommand{\pchisq}{\ensuremath{P\left(>\chisq \right)}}
\newcommand{\chisq}{\ensuremath{\chi^{\,2}}}
\newcommand{\chisqr}{\ensuremath{\chi^{\,2}_{r}}}
\newcommand\mysim{\mathord{\sim}}
\newcommand{\multifast}{{\tt MULTIFAST}}

\begin{document}

\title{KELT-19A\MakeLowercase{b}: A P$\sim$4.6 day hot jupiter transiting a likely A\MakeLowercase{m} star with a distant stellar companion}

\author{
Robert J. Siverd\altaffilmark{1},   
Karen A.\ Collins\altaffilmark{2,3}, 
George Zhou\altaffilmark{2,4},
Samuel N. Quinn\altaffilmark{2},
B. Scott Gaudi\altaffilmark{5},
Keivan G. Stassun\altaffilmark{6,7},
Marshall C. Johnson\altaffilmark{5},
Allyson Bieryla\altaffilmark{2},
David W. Latham\altaffilmark{2},
David R. Ciardi\altaffilmark{8},
Joseph E. Rodriguez\altaffilmark{2},
Kaloyan Penev\altaffilmark{9},
Marc Pinsonneault\altaffilmark{5},
Joshua Pepper\altaffilmark{10},
Jason D. Eastman\altaffilmark{2},
Howard Relles\altaffilmark{2},
John F. Kielkopf\altaffilmark{11},
Joao Gregorio\altaffilmark{12},
Thomas E. Oberst\altaffilmark{13},
Giulio Francesco Aldi\altaffilmark{14,15},  
Gilbert A. Esquerdo\altaffilmark{2},
Michael L. Calkins\altaffilmark{2},
Perry Berlind\altaffilmark{2},
Courtney D. Dressing\altaffilmark{16.17},
Rahul Patel\altaffilmark{18},
Daniel J. Stevens\altaffilmark{5},
Thomas G. Beatty\altaffilmark{19,20},
Michael B. Lund\altaffilmark{6},
Jonathan Labadie-Bartz\altaffilmark{10},
Rudolf B. Kuhn\altaffilmark{21,22},
Knicole D. Col\'{o}n\altaffilmark{23},
David James\altaffilmark{24},
Xinyu Yao\altaffilmark{10},
John A. Johnson\altaffilmark{2},
Jason T. Wright\altaffilmark{19,20},
Nate McCrady\altaffilmark{25},
Robert A. Wittenmyer\altaffilmark{26},
Samson A. Johnson\altaffilmark{5},
David H. Sliski\altaffilmark{27},
Eric L. N. Jensen\altaffilmark{28},
David H. Cohen\altaffilmark{28},
Kim K. McLeod\altaffilmark{29},
Matthew T. Penny\altaffilmark{5,30},
Michael D. Joner\altaffilmark{31},
Denise C. Stephens\altaffilmark{31},
Steven Villanueva Jr.\altaffilmark{5},
Roberto Zambelli\altaffilmark{32},
Christopher Stockdale\altaffilmark{33},
Phil Evans\altaffilmark{34},
Thiam-Guan Tan\altaffilmark{35},
Ivan A. Curtis\altaffilmark{36},
Phillip A. Reed\altaffilmark{37},
Mark Trueblood\altaffilmark{38},
Patricia Trueblood\altaffilmark{38}}

\altaffiltext{1}{Las Cumbres Observatory, 6740 Cortona Dr., Suite 102, Goleta, CA 93117, USA}
\altaffiltext{2}{Harvard-Smithsonian Center for Astrophysics, Cambridge, MA 02138, USA}
\altaffiltext{3}{Corresponding author; kcollins@cfa.harvard.edu}
\altaffiltext{4}{Hubble Fellow}
\altaffiltext{5}{Department of Astronomy, The Ohio State University, 140 West 18th Ave., Columbus, OH 43210, USA}
\altaffiltext{6}{Department of Physics and Astronomy, Vanderbilt University, Nashville, TN 37235, USA}
\altaffiltext{7}{Department of Physics, Fisk University, 1000 17th Avenue North, Nashville, TN 37208, USA}
\altaffiltext{8}{NASA Exoplanet Science Institute/Caltech, Pasadena, CA, USA}
\altaffiltext{9}{Department of Physics, The University of Texas at Dallas, 800 West Campbell Road, Richardson, TX 75080-3021 USA}
\altaffiltext{10}{Department of Physics, Lehigh University, 16 Memorial Drive East, Bethlehem, PA, 18015, USA}
\altaffiltext{11}{Department of Physics and Astronomy, University of Louisville, Louisville, KY 40292, USA}
\altaffiltext{12}{Atalaia Group \& CROW Observatory, Portalegre, Portugal}
\altaffiltext{13}{Department of Physics, Westminster College, New Wilmington, PA 16172, USA}
\altaffiltext{14}{Dipartimento di Fisica ``E. R. Caianiello'', Universit\`{a} di Salerno, Via Giovanni Paolo II 132, 84084 Fisciano (SA), Italy}
\altaffiltext{15}{Istituto Nazionale di Fisica Nucleare, Sezione di Napoli, 80126 Napoli, Italy} 
\altaffiltext{16}{NASA Sagan Fellow, Division of Geological \& Planetary Sciences, California Institute of Technology, Pasadena, CA 91125, USA}
\altaffiltext{17}{Department of Astronomy, University of California, Berkeley, CA 94720-3411, USA}
\altaffiltext{18}{IPAC, Mail Code 100-22, Caltech, 1200 E. California Blvd., Pasadena, CA 91125, USA}
\altaffiltext{19}{Department of Astronomy \& Astrophysics, The Pennsylvania State University, 525 Davey Lab, University Park, PA 16802, USA}
\altaffiltext{20}{Center for Exoplanets and Habitable Worlds, The Pennsylvania State University, 525 Davey Lab, University Park, PA 16802, USA}
\altaffiltext{21}{South African Astronomical Observatory, PO Box 9, Observatory, 7935 Cape Town, South Africa}
\altaffiltext{22}{Southern African Large Telescope, PO Box 9, Observatory, 7935 Cape Town, South Africa}
\altaffiltext{23}{NASA Goddard Space Flight Center, Greenbelt, MD 20771, USA}
\altaffiltext{24}{Astronomy Department, University of Washington, Box 351580, Seattle, WA 98195, USA}
\altaffiltext{25}{Department of Physics and Astronomy, University of Montana, Missoula, MT 59812, USA}
\altaffiltext{26}{University of Southern Queensland, Computational Engineering and Science Research Centre, Toowoomba, Queensland 4350, Australia}
\altaffiltext{27}{Department of Physics and Astronomy, University of Pennsylvania, Philadelphia, PA 19104, USA}
\altaffiltext{28}{Department of Physics and Astronomy, Swarthmore College, Swarthmore, PA 19081, USA}
\altaffiltext{29}{Department of Astronomy, Wellesley College, Wellesley, MA 02481, USA}
\altaffiltext{30}{Sagan Fellow}
\altaffiltext{31}{Department of Physics and Astronomy, Brigham Young University, Provo, UT 84602, USA}
\altaffiltext{32}{Societ\`a Astronomica Lunae, Castelnuovo Magra 19030, Italy}
\altaffiltext{33}{Hazelwood Observatory, Churchill, Victoria, Australia}
\altaffiltext{34}{El Sauce Observatory, Coquimbo Province, Chile}
\altaffiltext{35}{Perth Exoplanet Survey Telescope, Perth, Australia}
\altaffiltext{36}{ICO, Adelaide, South Australia}
\altaffiltext{37}{Department of Physical Sciences, Kutztown University, Kutztown, PA, 19530, USA)}
\altaffiltext{38}{Winer Observatory, Sonoita, AZ 85637, USA}

\shorttitle{KELT-19A\MakeLowercase{b}}

\begin{abstract}
We present the discovery of the giant planet KELT-19Ab, which transits the moderately bright ($V\sim 9.9$) A8V star TYC 764-1494-1 with an orbital period of 4.61\,d.  We confirm the planetary nature of the companion via a combination of radial velocities, which limit the mass to $\la 4.1\mj$ ($3\sigma$), and a clear Doppler tomography signal, which indicates a retrograde projected spin-orbit misalignment of $\lambda = -179.7^{+3.7}_{-3.8}$ degrees.  Global modeling indicates that the $\teff=7500 \pm 110$\,K host star has $\Mstar = 1.62^{+0.25}_{-0.20}\,M_\odot$ and $\Rstar = 1.83 \pm 0.10\,R_\odot$. The planet has a radius of $R_P=1.91 \pm 0.11\,\rj$ and receives a stellar insolation flux of $\sim 3.2\times 10^{9}\,{\rm erg\,s^{-1}\,cm^{-2}}$, leading to an inferred equilibrium temperature of $\teq \sim 1935$\,K assuming zero albedo and complete heat redistribution. With a $\vsini=84.8\pm 2.0$\kms, the host is relatively slowly rotating compared to other stars with similar effective temperatures, and it appears to be enhanced in metallic elements but deficient in calcium, suggesting that it is likely an Am star. KELT-19A would be the first detection of an Am host of a transiting planet of which we are aware.  Adaptive optics observations of the system reveal the existence of a companion with late G9V/early K1V spectral type at a projected separation of $\approx 160$\,AU. Radial velocity measurements indicate that this companion is bound. Most Am stars are known to have stellar companions, which are often invoked to explain the relatively slow rotation of the primary. In this case, the stellar companion is unlikely to have caused the tidal braking of the primary. However, it may have emplaced the transiting planetary companion via the Kozai-Lidov mechanism.  

\end{abstract}



\maketitle

\section{Introduction}

The Kilodegree Extremely Little Telescope (KELT; \citealt{Pepper:2003,Pepper:2007,Pepper:2012}) survey was originally designed to discover transiting planets orbiting bright $(8 \le V_{\mathrm{mag}} \le 11$) host stars. The scientific value and strategy behind that approach was described in detail in the introduction of the recent discovery of KELT-20b \citep{Lund:2017}\footnote{See also \citet{Talens:2017c} for the simultaneous discovery of the same planet, MASCARA-2b.}. In short, these bright systems are the most amenable to detailed follow-up characterization (i.e., transit spectroscopy, secondary eclipse spectroscopy, phase curve measurements, etc., \citealt{Winn:2010}). Because the KELT project did not actively start to vet candidates until 2011, many of the initial transit candidates had already been discovered by other collaborations (e.g., \citealt{Alonso:2004,McCullough:2006,Bakos:2007,CollierCameron:2007}).  

This fact, combined with a few additional, coincidental, and nearly-simultaneous occurrences, such as the confirmation of WASP-33b \citep{CollierCameron2010} via Doppler tomography (see \S \ref{sec:dopptom} for an overview of this technique), our somewhat fortuitous discovery of KELT-1b \citep{Siverd:2012}, and the `late entry' of KELT into the field of exoplanet discovery via transits, led us to pursue the discovery of transiting planets around hotter stars. This strategy has ultimately proven quite successful. In retrospect, the pursuit of hot stars was well-suited to the survey, both because KELT observes a larger fraction of hot stars than other ground-based transit surveys (due to Malmquist bias, see \citealt{Bieryla:2015}), but also because the reduction pipeline of the primary follow-up radial velocity vetting resource used by KELT, the Tillinghast Reflector Echelle Spectrograph (TRES) on the 1.5 m telescope at the Fred Lawrence Whipple Observatory, Mount Hopkins, Arizona, USA, was actively optimized to measure radial velocities of hot, rapidly-rotating stars \citep{Latham:2009}.

To date, this strategy of targeting hot stars has led to the discovery of four planets transiting A stars by the KELT survey: KELT-17b \citep{Zhou:2017}, KELT-9b \citep{Gaudi:2017}, KELT-20b/MASCARA-2b  \citep{Lund:2017,Talens:2017c}, and KELT-19Ab, the planet announced here. Additionally, there are four planets known to transit A stars discovered by other collaborations: WASP-33b \citep{CollierCameron2010}, Kepler-13Ab \citep{Shporer:2011}, HAT-P-57b \citep{Hartman:2015}, and MASCARA-1b \citep{Talens:2017b}. 

As discussed in previous KELT planet discovery papers, rapidly rotating, hot stars above the Kraft break \citep{Kraft:1967} pose unique challenges but provide unique opportunities. Transiting planets orbiting these stars are difficult to confirm via Doppler reflex motion, but on the other hand are amenable to Doppler tomography due to the large $\vsini$ of their hosts.   

It is also the case that A stars have a remarkable diversity in their properties, partially due to the fact that their outer envelopes are primarily radiative, but exhibit extremely thin helium and hydrogen convective layers at the very outer edges of their atmospheres. In particular, the thin surface convection zones and very low mass loss rates of A stars lead to very efficient gravitational settling of some elements, similar to (although not as extreme as) the settling exhibited in white dwarfs. This results in weaker spectral lines of those elements relative to what would be expected of a star of similar temperature, and not indicative of an actual global underabundance of those elements. Similarly, because the convective zones are so thin, partially ionized elements with large radiative cross sections below the convective zone can exhibit radiative levitation. This may lead to stronger lines which may be interpreted as large selective overabundances in certain elements (see, e.g., \citealt{Richer:2000}). Indeed, it is even possible to have an element experience both gravitational settling and radiative levitation in different layers, creating a zone within an A star where that element is highly concentrated. In the case of iron, this effect may be severe enough to induce convective mixing that can impact surface abundances \citep{Richard:2001}.

In general, thinner surface convection zones that are lower in density experience gravitational settling at a faster rate, and are more susceptible to radiative levitation. In normal A stars, there are thin hydrogen and helium ionization zones that are very close to each other, which through overshoot behave as a single deeper mixed layer. However, if the helium ionization zone is driven much deeper and no longer in causal contact with the hydrogen ionization zone, even more extreme abundance changes may be apparent, since the hydrogen ionization zone by itself is isolated and very shallow.  

The net result is that determining the global metal abundances for A stars can be extremely difficult. Abundances determined by atmospheric spectroscopy may have very little to do with the global metallic abundance of the star. A particularly notable example is the metallic-line Am stars \citep{Titus:1940}, which, although they have hydrogen lines consistent with the effective temperatures of late A stars, also have metallic lines of heavier elements with strengths expected for cooler F stars, and lines of lighter elements consistent with hotter A stars. These Am stars are generally more slowly rotating  \citep{Abt:1995} than chemically normal stars with the same effective temperatures, likely due to a competition between elemental segregation and rotational mixing. The net result is that surface abundance anomalies can be enhanced in some elements and suppressed in others for Am stars \citep{Abt:1995}. Empirically, stars with rotational speeds above $\sim 150~$\kms\ are chemically ``normal'' and it appears that mixing overcomes the settling described above. Virtually all slower rotators (including KELT-19A) are measured to be chemically peculiar, although there may well be exceptions. For example, a very young slow rotator might not yet have had time to develop unusual abundance patterns. Empirically, most slowly rotating Am stars are also in binaries \citep{Abt:1985}, as is the case for KELT-19A (see \S \ref{sec:ao} and \S \ref{sec:absRV}). This may be due to tidal braking of the A star, although in the case of KELT-19, the stellar companion is too distant for such tidal braking to be effective. 

\section{Discovery and Follow-Up Observations}

We provide a brief summary of the KELT survey data reduction process and present the results in \S \ref{sec:keltobs}. \S \ref{sec:phot} presents our ground-based time-series follow-up photometric observations, \S \ref{sec:ao} presents our high contrast adaptive optics imaging, and \S \ref{sec:spectroscopy} presents our spectroscopic follow-up observations.

\subsection{KELT Observations and Photometry}\label{sec:keltobs}

\begin{figure}
\centering 
\includegraphics[width=1.0\linewidth, angle=0, trim = 0 0 0 0]{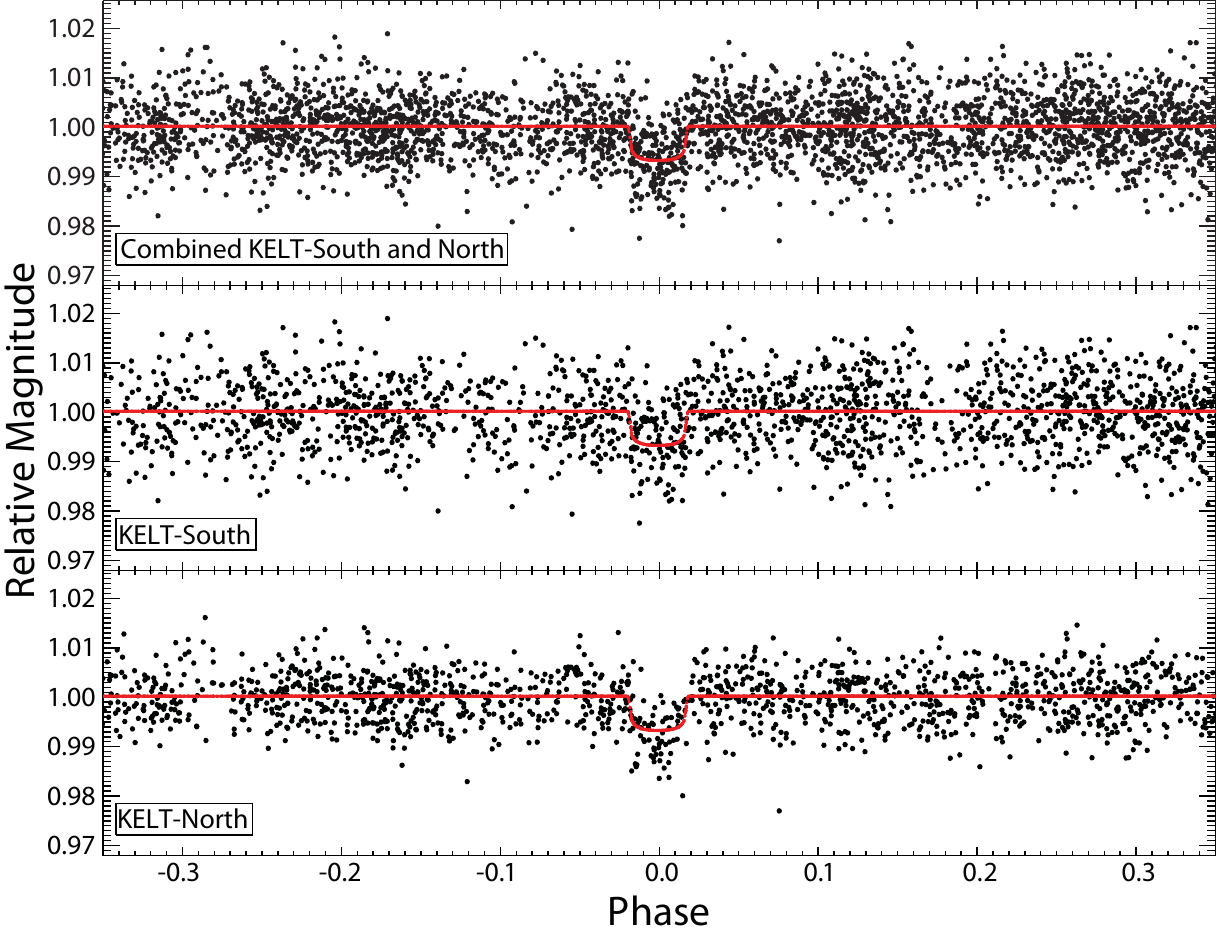}
\caption{\footnotesize The combined KELT-South and KELT-North light curve (top), the KELT-South light curve only (middle), and the KELT-North discovery light curve (bottom) for KELT-19Ab.  Each has been phase-folded to the discovery period of 4.6117449 days. The red line corresponds to an EXOFAST model of the combined light curve. (Supplemental data for this figure are available in the online journal.)}
\label{fig:DiscoveryLC}
\end{figure}

KELT-19Ab is located in a field that is monitored by both KELT telescopes, centered on $\alpha =$ 07$^{h}$ 39$^{m}$ 36$^{s}$, $\delta =$ $+03\degr$ 00$\arcmin$ 00$\arcsec$ (J2000). This field is labeled internally as KELT-South field 06 (KS06) and KELT-North field 14 (KN14). The reduction and candidate selection process for KELT-South and KELT-North are described in detail in \citet{Kuhn:2016} and \citet{Siverd:2012}, respectively. From our analysis of 2636 images from KS06 (UT 2010 March 02 to 2013 May 10) and 2092 images from KN14 (UT 2011 October 11 to UT 2013 March 26), KJ06C009789 (KELT-19Ab) was identified as a top candidate. Figure \ref{fig:DiscoveryLC} shows the combined KELT-South and KELT-North light curve (top), the KELT-South light curve only (middle), and the KELT-North light curve (bottom) for KELT-19Ab. KELT-19 (BD+07 1721) is located at $\alpha =$ 07$^{h}$ 26$^{m}$ 02$\fs$2895, $\delta =$ +07$\degr$ 36$\arcmin$ 56$\farcs$834 (J2000). This is the second planet discovered through a combination of KELT-South and KELT-North observations, KELT-17b being the first one \citep{Zhou:2016}. 

\subsection{Photometric Time-series Follow-up} \label{sec:phot}

\begin{figure}
\vspace{0in}
\includegraphics[width=\linewidth,angle=0,trim=0 0 0 0,clip=true]{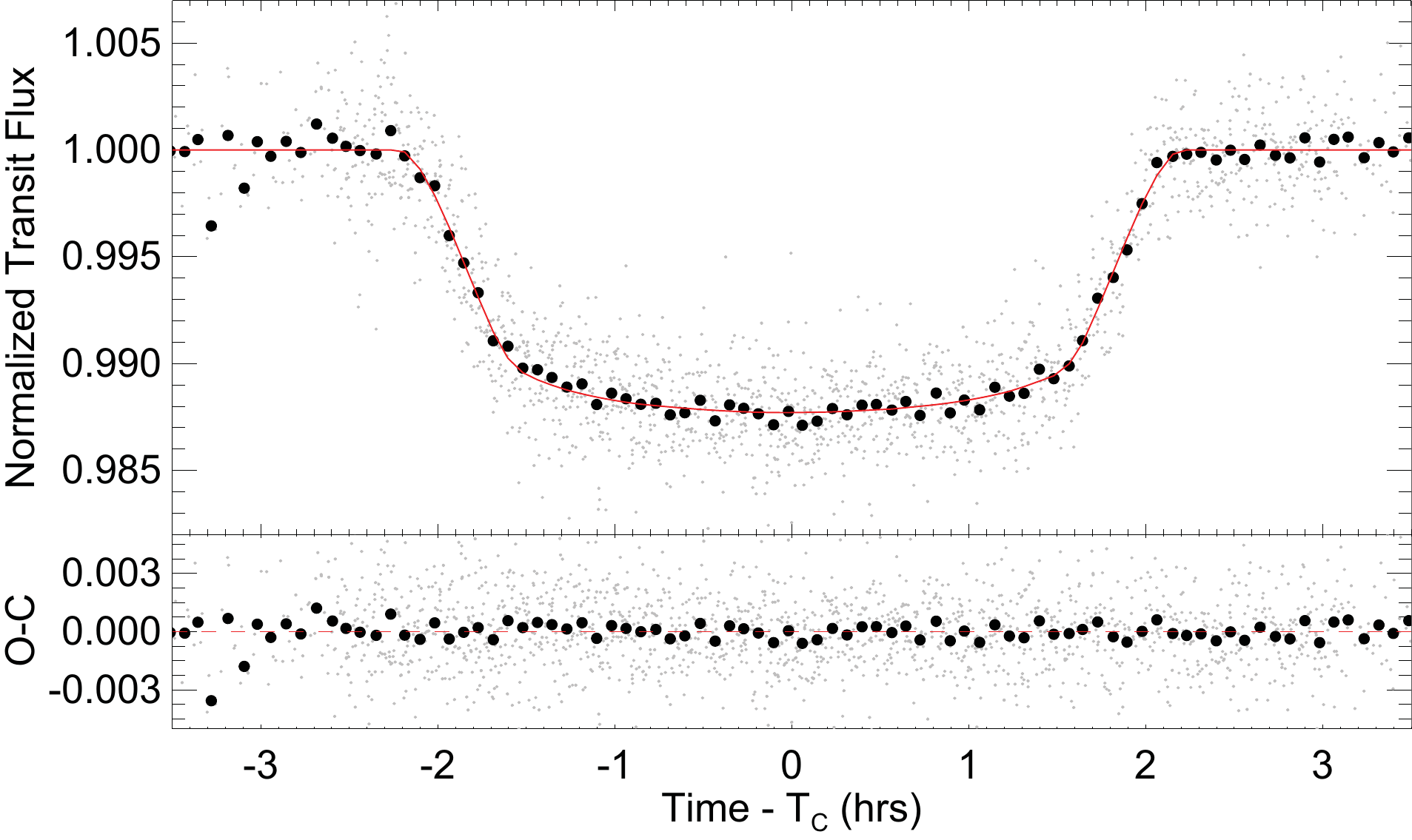}
\caption{\footnotesize All detrended follow-up transits combined (small light-gray points) and binned in 5 minute intervals (large black points) and overplotted with the best fit global model (solid red line). The model shows the average limb darkening weighted by the number of transits in each band. The model residuals are shown in the bottom panel. The binned light curve data are not used in the analysis and are presented here to illustrate the overall statistical power of the follow-up photometry. }
\label{fig:binned_lcs} 
\end{figure}

The KELT collaboration includes a world-wide team of ground-based follow-up observers known as the KELT Follow-Up Network (KELT-FUN). KELT-FUN currently includes members from $\approx$ 60 institutions. The KELT-FUN team acquired follow-up time-series photometry of KELT-19Ab transits to better determine the system parameters and to check for transit false positives. We used the {\tt Tapir} software package \citep{Jensen:2013} to schedule follow-up observations. We obtained six full and three partial transits in multiple bandpasses from $g$ to $z$ between February 2015 and December 2016. Figure \ref{fig:primary_lcs} shows all the transit follow-up light curves assembled. A summary of the follow-up photometric observations is shown in Table \ref{tbl:photobs}. We find consistent $R_{\rm P}/R_\star$ ratios in all light curves across the optical bands, helping to rule out false positives due to blended eclipsing binaries. Figure \ref{fig:binned_lcs} shows all transit followup
light curves from Figure \ref{fig:primary_lcs} combined and binned in 5 minute intervals. This combined and binned light curve is not used for analysis, but rather to show the best combined behavior of the transit.

All photometric follow-up observations were reduced with the {\tt AstroImageJ} ({\tt AIJ}) software package\footnote{http://www.astro.louisville.edu/software/astroimagej/} \citep{Collins:2017}. We were careful to ensure that all observatory computers were referenced either through a network connection to a stratum 1 timing source or to a GPS stratum 1 timing source, and that all quoted mid-exposure times were properly reported in barycentric Julian dates at mid-exposure (\bjdtdb; \citealt{Eastman:2010}). 

\subsubsection{KeplerCam}
We observed an $i$-band transit ingress from KeplerCam on the 1.2\,m telescope at the Fred Lawrence Whipple Observatory (FLWO) on UT 2015 February 20. KeplerCam has a single $4{\rm K} \times 4{\rm K}$ Fairchild CCD 486 with an image scale of $0\farcs366$ pixel$^{-1}$ and a field of view of $23\farcm1 \times 23\farcm1$.

\subsubsection{WCO} 
We observed an $r$-band transit egress from the Westminster College Observatory (WCO) on UT 2015 March 06. The observations were conducted from a 0.35\,m f/11 Celestron C14 Schmidt-Cassegrain telescope equipped with an SBIG STL-6303E CCD with a $3{\rm K}\times2{\rm K}$ array of 9\,$\mu$m pixels. The resulting images have a $24\arcmin\times16\arcmin$ field of view and 1$\farcs$4 pixel$^{-1}$ image scale at $3\times3$ pixel binning.

\subsubsection{Salerno}
We observed an $R$-band transit ingress on UT 2015 March 19 from the Salerno University Observatory in Fisciano Salerno, Italy. The observing setup consists of a 0.35\,m Celestron C14 SCT and an SBIG ST2000XM $1600\times1200$ CCD, yielding an image scale of $0\farcs54$ pixel$^{-1}$.

\subsubsection{MINERVA}

We observed a full transit simultaneously in the Sloan $r$-, $i$-, and $z$-bands using three of the MINERVA Project telescopes \citep{Swift:2015} on the night of UT 2016 January 18. MINERVA uses four 0.7\,m PlaneWave CDK-700 telescopes that are located on Mt. Hopkins, AZ, at FLWO. While the four telescopes are normally used to feed a single spectrograph, we used three MINERVA telescopes in their photometric imaging mode for the KELT-19 observations. The telescopes were equipped with Andor iKON-L $2048\times2048$ cameras, which gave a field of view of $20\farcm9 \times 20\farcm9$ and a plate scale of $0\farcs6$ pixel$^{-1}$.The MINERVA telescope conducting the $r$-band observations experienced a $2\farcm6$ tracking jump during the time of egress. The resulting change in the position of the field on the detector produces a relatively large change in the baseline level of the light curve just after the beginning of egress. Furthermore, because of imperfect flat-field images, the baseline offset differs by $\sim\pm1$ percent depending on the set of comp stars selected. The different baseline offsets produce transit center times that differ by $\sim\pm8$~minutes, even with detrending parameters included in the model to attempt to compensate for the baseline offset. Because of the unreliable detrending results, and the fact that we simultaneously observed four additional light curves on UT 2016 January 18, the $r$-band light curve is not included in the analysis to avoid the potential of improperly biasing the linear ephemeris derived from our global modelling effort (see \S \ref{sec:ttvs}).

\subsubsection{MVRC}

We observed a full transit from the Manner-Vanderbilt Ritchey-Chr\'{e}tien (MVRC) telescope located at the Mt. Lemmon summit of Steward Observatory, AZ, on UT 2016 January 18. Exposures were taken in alternating $g$- and $i$-band filters yielding pseudo-simultaneous observations in the two filters. The observations employed a 0.6 m f/8 RC Optical Systems Ritchey-Chr\'{e}tien telescope and an SBIG STX-16803 CCD with a $4{\rm K}\times4{\rm K}$ array of 9\,$\mu$m pixels, yielding a $26\farcm6 \times 26\farcm6$ field of view and $0\farcs39$ pixel$^{-1}$ image scale. 

\subsubsection{CROW} 
We observed a full $I$-band transit from Canela’s Robotic Observatory (CROW) in Portalegre, Portugal on UT 2016 December 05. The observatory is equipped with a 0.3\,m Schmidt-Cassegrain telescope and a KAF-3200E CCD, having a 30$\arcmin\times20\arcmin$ field of view and a pixel scale of $0\farcs84$ pixel$^{-1}$.

\begin{figure}
\begin{center}
\includegraphics[width=\linewidth,angle=0,trim=0 0 0 0,clip=true]{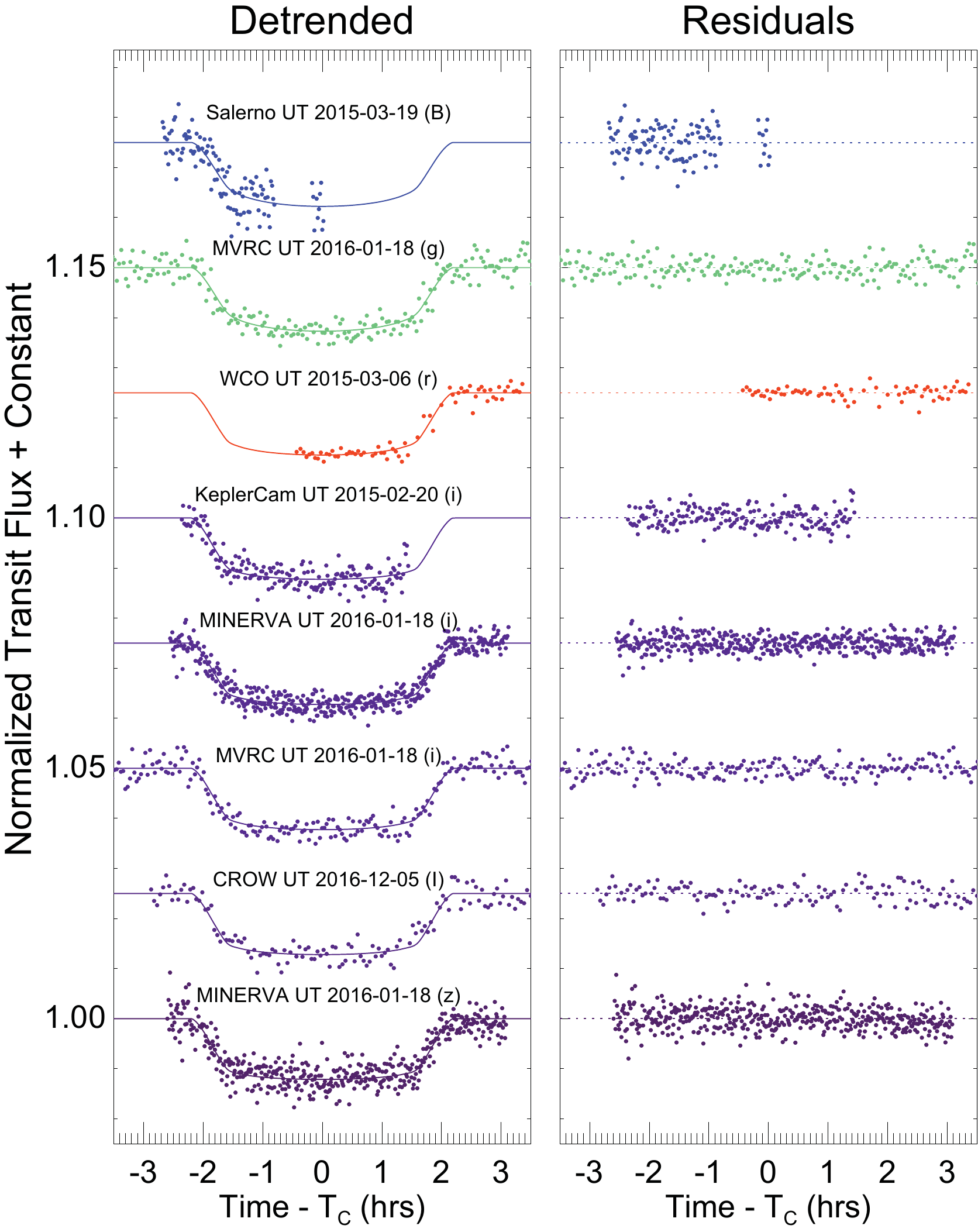}
\caption{Follow-up transit photometry of KELT-19. {\it Left Panel:} Detrended transit light curves arbitrarily shifted on the vertical axis for clarity. The overplotted solid lines are the best fit transit model from the adopted global fit documented in Table \ref{tab:parameters}. {\it Right Panel:} The transit model residuals. The labels are as follows:
Salerno=Salerno University Observatory 0.35\,m telescope;
MVRC=Manner-Vanderbilt 0.6\,m RCOS Telescope;
WCO=Westminster College Observatory 0.35\,m telescope;
KeplerCam=1.2\,m telescope at FLWO;
MINERVA=MINiature Exoplanet Radial Velocity Array of 0.7 m telescopes;
CROW=Canela's Robotic Observatory 0.3 m LX200 Telescope.
(Supplemental data for this figure are available in the online journal.)
\label{fig:primary_lcs}}
\end{center}
\end{figure}

\begin{table}
\begin{center}
\caption{Summary of Photometric Observations\label{tbl:photobs}}
{\setlength{\tabcolsep}{0.30em}
\begin{tabular}{lcccccccc}
\tableline
\multicolumn{1}{l}{Telescope} & \multicolumn{1}{c}{UT} & \multicolumn{1}{c}{\#} & \multicolumn{1}{c}{Fil-} & \multicolumn{1}{c}{Cyc\tablenotemark{a}} & \multicolumn{1}{c}{RMS\tablenotemark{b}} & \multicolumn{1}{c}{PNR\tablenotemark{c}} & \multicolumn{1}{c}{Error\tablenotemark{d}} & \multicolumn{1}{c}{Detrend} \\ 
\multicolumn{1}{l}{ } & \multicolumn{1}{c}{Date} & \multicolumn{1}{c}{Obs} & \multicolumn{1}{c}{ter} & \multicolumn{1}{c}{(sec)} & \multicolumn{1}{c}{{\tiny ($10^{\text{-}3}$)}}  & \multicolumn{1}{c}{\tiny $(\frac{10^{\text{-}3}}{min.})$} & \multicolumn{1}{c}{Scale} & \multicolumn{1}{c}{Data}\\
\vspace{-0.1in}\\
\tableline
\vspace{-0.1in}\\
KeplerCam  & 2015-02-20  & 185 & i  &  74 &  1.9 &  2.1  & 0.944 & AM\\
WCO        & 2015-03-06  &  67 & r  & 220 &  1.9 &  3.6  & 1.046 & TM\\ 
Salerno    & 2015-03-19  & 110 & B  &  68 &  3.1 &  3.3  & 1.429 & AM\\
MVRC       & 2016-01-18  & 236 & g  &  53 &  2.2 &  2.1  & 2.791 & AM,SK\\
MVRC       & 2016-01-18  & 236 & i  &  83 &  1.8 &  2.1  & 2.435 & AM\\
MINERVA    & 2016-01-18  & 446 & i  &  46 &  1.5 &  1.3  & 1.874 & AM,FW\\
MINERVA    & 2016-01-18  & 444 & z  &  46 &  2.1 &  1.8  & 1.774 & AM\\
CROW       & 2016-12-05  & 128 & I  & 186 &  1.7 &  3.0  & 1.827 & MF,FW\\
\tableline
\end{tabular}}
\tablecomments{
See Figure \ref{fig:primary_lcs} for a description of the telescope naming convention; 
AM=airmass;
TM=time;
SK=sky background;
FW=average FWHM in image;
MF=baseline offset at meridian flip.
}
\vspace{-10pt}
\tablenotetext{1}{Cycle time in seconds, calculated as the mean of exposure time plus dead time during periods of back-to-back exposures.}
\tablenotetext{2}{RMS of residuals from the best fit model in units of $10^{-3}$ .}
\tablenotetext{3}{Photometric noise rate in units of $10^{-3}$ minute$^{-1}$, calculated as RMS/$\sqrt{\Gamma}$, where RMS is the scatter in the light curve residuals and $\Gamma$ is the mean number of cycles (exposure time and dead time) per minute during periods of back-to-back exposures (adapted from \citealt{Fulton:2011}).}
\tablenotetext{4}{Error scaling factor determined by \multifast (see \S \ref{sec:globalfit}).}
\end{center}
\end{table}

\subsection{High-Contrast Imaging}\label{sec:ao}

KELT-19 was observed on the night of UT 2016 December 18 at Palomar Observatory with the 200$^{\prime\prime}$ Hale Telescope using the near-infrared adaptive optics (AO) system P3K and the infrared camera PHARO \citep{Hayward:2001}.  PHARO has a pixel scale of $0\farcs025$ pixel$^{-1}$ and a full field of view of approximately $25\arcsec$. The data were obtained with a narrow-band $Br$-$\gamma$ filter $(\lambda_o = 2.18; \Delta\lambda = 0.03\,\mu$m ) and a standard $J$-band filter $(\lambda_o = 1.246; \Delta\lambda = 0.162\,\mu$m).

The AO data were obtained in a 5-point quincunx dither pattern with each dither position separated by $4\arcsec$.  Each dither position is observed 3 times, each offset from the previous image by $0\farcs5$ for a total of 15 frames; the integration time per frame was 17 seconds in both the $Br$-$\gamma$ and $J$ filters. We use the dithered images to remove sky background and dark current, and then align, flat-field, and stack the individual images. The PHARO AO data have a resolution of $0\farcs11$ and $0\farcs25$ (FWHM) in the $Br$-$\gamma$ and $J$ filters, respectively.

\begin{figure}[!hb]
\begin{center}
\includegraphics[width=1.0\linewidth,trim=20mm 5mm 10mm 10mm,clip]{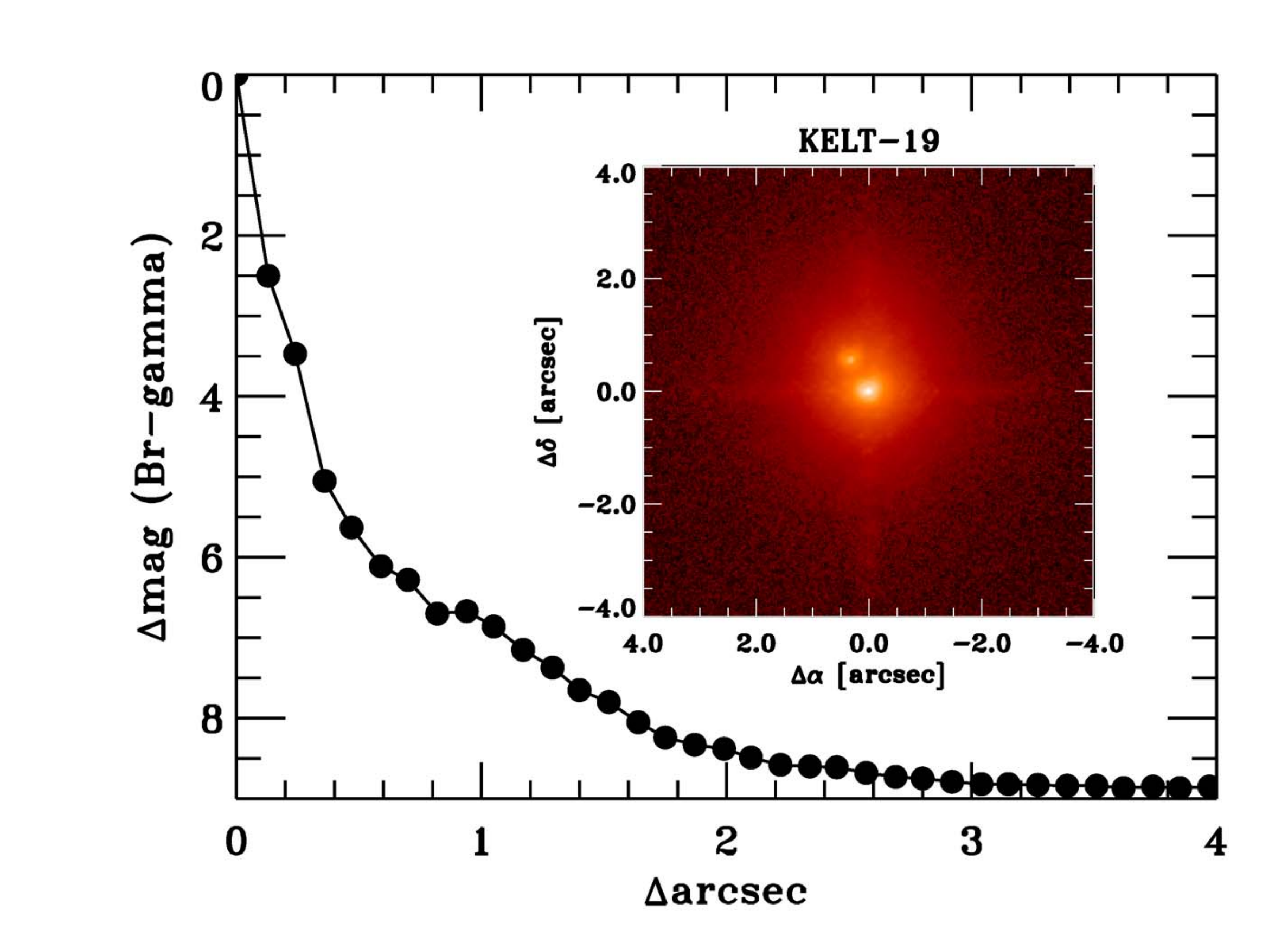}
\caption{Contrast sensitivity and inset image of KELT-19 in $Br$-$\gamma$ as observed with the Palomar Observatory Hale Telescope adaptive optics system; the secondary companion is clearly detected.  The $5\sigma$ contrast limit in $\Delta$magnitude is plotted against angular separation in arcseconds. The slight decrease in sensitivity near $1\arcsec$ is caused by an increase in the relative brightness of the diffraction spikes in comparison to the smoothly declining point spread 
function of the target.} \label{fig:ao_contrast}
\end{center}
\end{figure}

The sensitivities of the final combined AO image were determined by injecting simulated sources azimuthally around KELT-19A every $45\degr$ at separations of integer multiples of the central source's FWHM \citep{Furlan:2017}. The brightness of each injected source was scaled until standard aperture photometry detected it with $5\sigma$ significance. The resulting brightness of the injected sources relative to KELT-19A set the contrast limits at that injection location. The $5\sigma$ limit at each separation was determined from the average of all of the determined limits at that separation. The contrast sensitivity curve shown in Figure \ref{fig:ao_contrast} represents the $5\sigma$ limits of the imaging data in $\Delta$magnitude versus angular separation in arcseconds.  The slight decrease in sensitivity near $1\arcsec$ is caused by an increase in the relative brightness of the diffraction spikes in comparison to the smoothly declining point spread function of the target.

For KELT-19, a nearby stellar companion was detected in both the $Br$-$\gamma$ and $J$ filters. The presence of the blended companion must be taken into account to obtain the correct transit depth and planetary radius (e.g. \citealt{Ciardi:2015}).  The companion separation was measured from the $Br$-$\gamma$ image and found to be $\Delta\alpha = 0\farcs32 \pm 0\farcs02$ and $\Delta\delta = 0\farcs55 \pm 0\farcs02$, which is a projected separation of $0\farcs64\pm0\farcs03$ at a position angle of $30.2\pm2.5$ degrees.  At a distance of $255\pm15$\,pc (see \S\ref{sec:SED}), the companion has a projected separation from the primary star of $\approx 160$ AU. The positional offset uncertainties between the two stars are based upon the uncertainties in the positional fit to the centroids of the point spread functions of the stars and is approximately 0.1 of a pixel corresponding to 2.5 mas.  No distortion map was applied to the images; however, the optical distortion is 0.4\% or less in the narrow camera mode for PHARO \citep{Hayward:2001}.

The stars have blended 2MASS magnitudes of $J = 9.343 \pm 0.026$ mag and $K_s = 9.196 \pm 0.023$ mag.  The stars have measured magnitude differences of $\Delta J = 2.50 \pm 0.06$ mag and $\Delta K_s = 2.045 \pm 0.03$ mag; the $J$-band differential measurement is less certain because of the poor AO correction in that filter on the night of the observations. $Br$-$\gamma$ has a central wavelength that is sufficiently close to $Ks$ to enable the deblending of the 2MASS magnitudes into the two components.  The primary star has deblended (real) apparent magnitudes of $J_1 = 9.45 \pm 0.03$ mag and $Ks_1 = 9.35 \pm 0.02$ mag, corresponding to a color of $(J-K_s)_1 = 0.10\pm 0.04$ mag; the companion star has deblended (real) apparent magnitudes of $J_2 = 11.95 \pm0.06$ mag and $Ks_2 = 11.40 \pm 0.03$ mag, corresponding to a color $(J-K_s)_2 = 0.55 \pm 0.07$ mag.  The uncertainties in the stellar colors are dominated by the uncertainty in the $J$-$band$ measurement. Using the \citet{Casagrande:2010} relations, the colors give $\teff=7190^{+270}_{-250}$\,K for the primary and $\teff=5030^{+260}_{-240}$\,K for the companion, which are consistent with the effective temperatures derived from the SED analysis in \S \ref{sec:SED} and the spectral analysis in \S \ref{sec:stellarPars}.

\subsection{Spectroscopic Follow-up} \label{sec:spectroscopy}

\subsubsection{TRES at FLWO}

To constrain the planet mass and enable eventual Doppler tomographic (DT) detection of KELT-19Ab, we obtained a total of 60 spectroscopic observations of the host star with the Tillinghast Reflector Echelle Spectrograph (TRES) on the 1.5\,m telescope at the Fred Lawrence Whipple Observatory, Arizona, USA. Each spectrum delivered by TRES has a spectral resolution of $\lambda / \Delta \lambda = 44000$ over the wavelength range of $3900-9100$\,\AA\ in 51 \'echelle orders. A total of 7 observations were obtained during the out-of-transit portions of the planet's orbit to constrain its mass (\S \ref{sec:absRV}). Two spectroscopic transits were observed, on 2016-02-24 and 2016-11-08, for the Doppler tomographic analysis. The observations on 2016-02-24 were plagued by bad weather, and were discarded. The transit sequence obtained on 2016-11-08, totaling 24 spectra, successfully revealed the planetary transit, and were used in the analysis described in \S \ref{sec:dopptom}.

\subsubsection{HJST at McDonald}

To provide additional constraints on the planet mass, we obtained 14 spectra of KELT-19 covering the entire orbital phase with the 2.7 m Harlan J.\,Smith Telescope (HJST) at McDonald Observatory and the Robert G.\,Tull Coud\'e spectrograph \citep{Tull:1995} in its TS23 configuration. This is a cross-dispersed \'echelle spectrograph with a resolving power of $R=60,000$ and coverage from 3570 to 10200 \AA\,(complete below 5691 \AA) over 58 orders. The first two spectra (from 2016 October) have exposure lengths of $\sim375$ seconds, while the last twelve (from 2016 December) have 1200 second exposure lengths. 

\subsubsection{Radial Velocities}\label{sec:absRV}

\begin{figure}
\vspace{0in}
\includegraphics[width=\linewidth,angle=0,trim=0 0 0 0,clip=true]{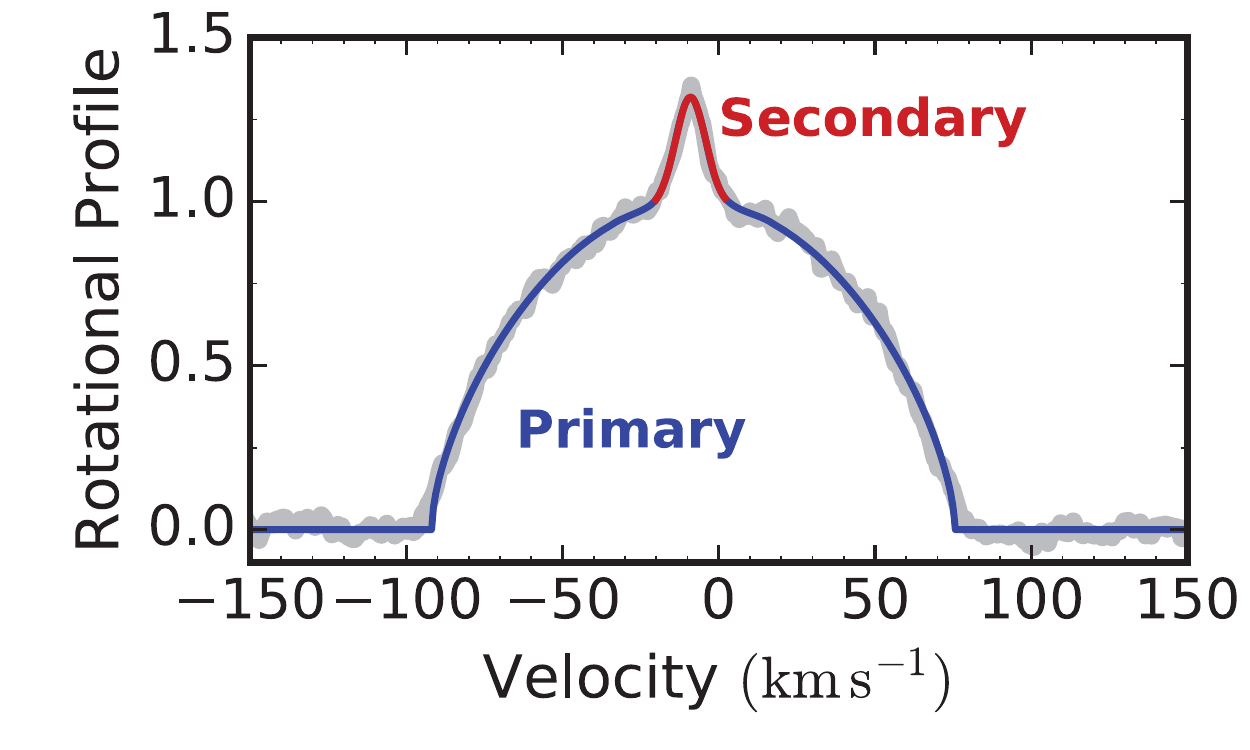}
\caption{\footnotesize An example broadening kernel of KELT-19 (gray line), as observed by TRES, showing the spectroscopic binary nature of the system. We fit both spectroscopic components simultaneously to obtain the radial velocities of both stellar components; the best fit profiles for the primary and companion are shown in blue and red, respectively. }
\label{fig:lsdfit} 
\end{figure}

The nearby stellar companion (see \S \ref{sec:ao}) is blended with the primary in our spectroscopic observations. Because of the resulting composite spectra, our radial velocity analysis is somewhat modified from previous KELT papers. For each observation, we derived a line broadening kernel via a least-squares deconvolution \citep[following][]{Donati1997,CollierCameron2010}, from which both spectroscopic components can be identified (Figure~\ref{fig:lsdfit}). To derive radial velocities and rotational broadening parameters, we fit for the two spectroscopic components simultaneously across all available out-of-transit spectra, allowing for independent radial velocities of the two components, whilst requiring all observations to have the same velocity broadening parameters. The TRES and HJST observations were fit independently, since they are subjected to different instrumental broadening, and the broadening profiles were derived from different spectral wavelength regions. 

From the simultaneous fit, we find that the out-of-transit broadening profile can best be described by a rapidly rotating primary star and a faint, slowly rotating, secondary star. The primary component has a rotational broadening velocity of $\vsini = 84.1 \pm 2.1$\kms\ and a combined macro- and microturbulent broadening of $3.4 \pm 2.0$\kms. The effect of instrumental broadening is taken into account separately in the global modelling. The secondary component has a line broadening velocity of $8.23\pm 0.11$\kms, which includes the influence of instrumental (6.8\kms), rotational, and turbulent sources. We find a flux ratio of $F_{\mathrm{B}}/F_{\mathrm{A}} = 0.0270 \pm 0.0034$ to the total light of the system over the wavelength range $5200\pm150$\,\AA. This flux ratio is consistent with the AO observations of the spatially separated companion, and with the interpretation that the secondary companion is a G-dwarf associated with the system (\S \ref{sec:SED}).

We estimate the absolute center of mass radial velocity for KELT-19 from the Mg b region of our TRES spectra. We examined the mean of (1) all velocities, (2) the out-of-transit velocities, and (3) the high SNR velocities and concluded that the best nominal value and uncertainty representing the absolute radial velocity of the KELT-19 system is $-7.9\pm 0.5$\kms. The absolute RV was then adjusted to the International Astronomical Union (IAU) Radial Velocity Standard Star system via a correction of $-0.62$\kms\ resulting in a final value of RV$_{\rm IAU}=-8.5\pm0.5$\kms. The correction primarily adjusts for the gravitational red-shift, which is not included in the library of synthetic template spectra.

\begin{figure}
\vspace{0in}
\includegraphics[width=\linewidth,angle=0,trim=0 0 0 0,clip=true]{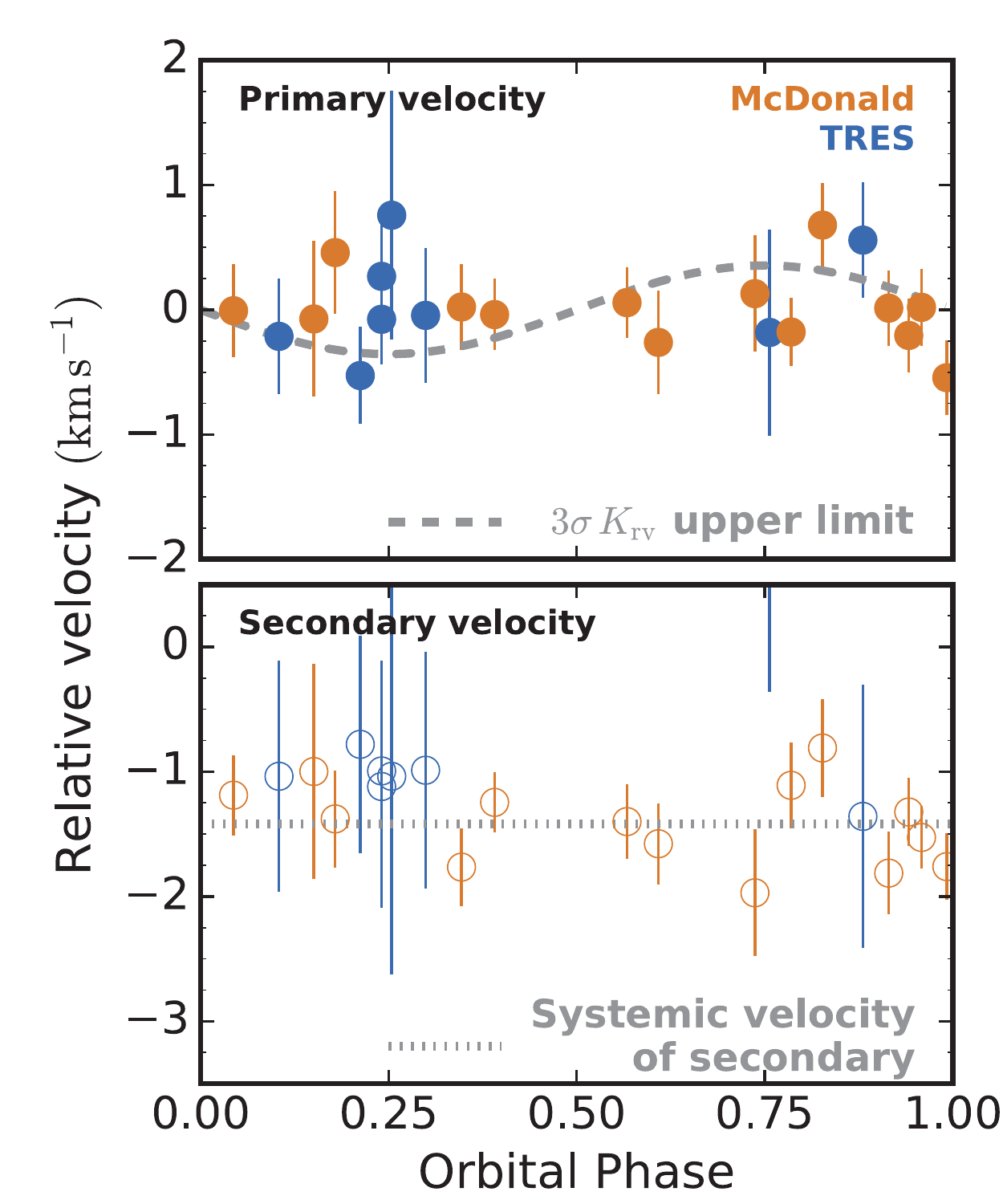}
\caption{\footnotesize Radial velocities of the two stellar components in the KELT-19 system, phase folded to the transit period. We can place a $3\sigma$ upper limit on the radial velocity semi-amplitude $K$ of 0.35\kms, confirming that the transiting companion is of planetary mass. The velocity of the stellar companion is constrained to be $<2.31$\kms\ at $1\sigma$, and $<7.50$\kms\ at $3\sigma$. The systemic velocity of the companion is similar to that of the primary, consistent with the interpretation that they are physically associated. The primary velocities are plotted in the top panel, secondary velocities in the bottom panel. The systemic velocity of the primary has been subtracted for all measurements. The TRES velocities are plotted in blue, McDonald in orange. }
\label{fig:rvlsd} 
\end{figure}

The TRES and HJST out-of-transit velocities are shown in Figure~\ref{fig:rvlsd}, and presented in Table~\ref{tab:rv1}. As discussed in \S \ref{sec:globalfit}, the radial velocity semi-amplitude of the primary can be constrained to be $<0.35$\kms, confirming that the transiting companion is of planetary mass. We also confirmed that the velocity of the stellar companion is not varying, and is constrained to be $<2.31$\kms\ at $1\sigma$, and $<7.50$\kms\ at $3\sigma$. We note that the biggest hurdle to obtaining precise radial velocities for KELT-19A is its rapid rotational velocity. In comparison, we often reach $\sim 10$\,m\,s$^{-1}$ precision for slowly rotating non-active stars with TRES \citep{Quinn:2014}. The systemic velocity of the primary ($-8.5\pm0.5$\kms) is consistent with that of the companion ($-9.4\pm1.0$\kms), which we interpret as KELT-19A and KELT-19B being bound. Assuming a 0.5\msun\ bound companion in a circular, nearly edge-on orbit with radius 160\,AU, KELT-19B would cause a maximum acceleration of KELT-19A (at conjunction or opposition) of $\sim 4$\,m\,s$^{-1}$\,yr$^{-1}$. Given the current relatively low RV precision due to the rapid rotation of the primary, it is not surprising that an RV trend is not detected in the current data, and furthermore would not be detected for the foreseeable future. However, under the same assumptions, KELT-19A would cause a maximum acceleration of KELT-19B of $\sim 12$\,m\,s$^{-1}$\,yr$^{-1}$, which might be detectable after several years with radial velocity instruments that can achieve precisions of a few m\,s$^{-1}$ for a $\mathrm{J}=12$ mag star, given the relatively low $\vsini$ of $\sim 4$\kms\ of the secondary.  

\subsubsection{Doppler Tomographic Observations}\label{sec:dopptom}

As a star rotates, one hemisphere moves toward the observer relative to the integrated stellar radial velocity, which produces light with a blue-shifted spectrum. The other hemisphere moves away from the observer, producing light with a red-shifted spectrum. In total, this produces rotationally broadened spectral lines. As a planet transits a star, differing blue- and/or red-shifted stellar spectral components are obscured by the shadow of the planet on the star. The planet shadow thus produces a spectral line profile distortion that varies in velocity space (except for the case of a polar orbit) as the transit progresses from ingress to egress. The measurement of the motion of the distortion can be modeled to reveal the system's spin-orbit misalignment, $\lambda$, and the impact parameter, $b$, of the planet's orbit relative to the stellar disk. See \citet{Johnson:2014} for a more technical description. 

\begin{figure}
\begin{center}
\includegraphics[width=1.0\linewidth,trim=0mm 0mm 0mm 0mm,clip]{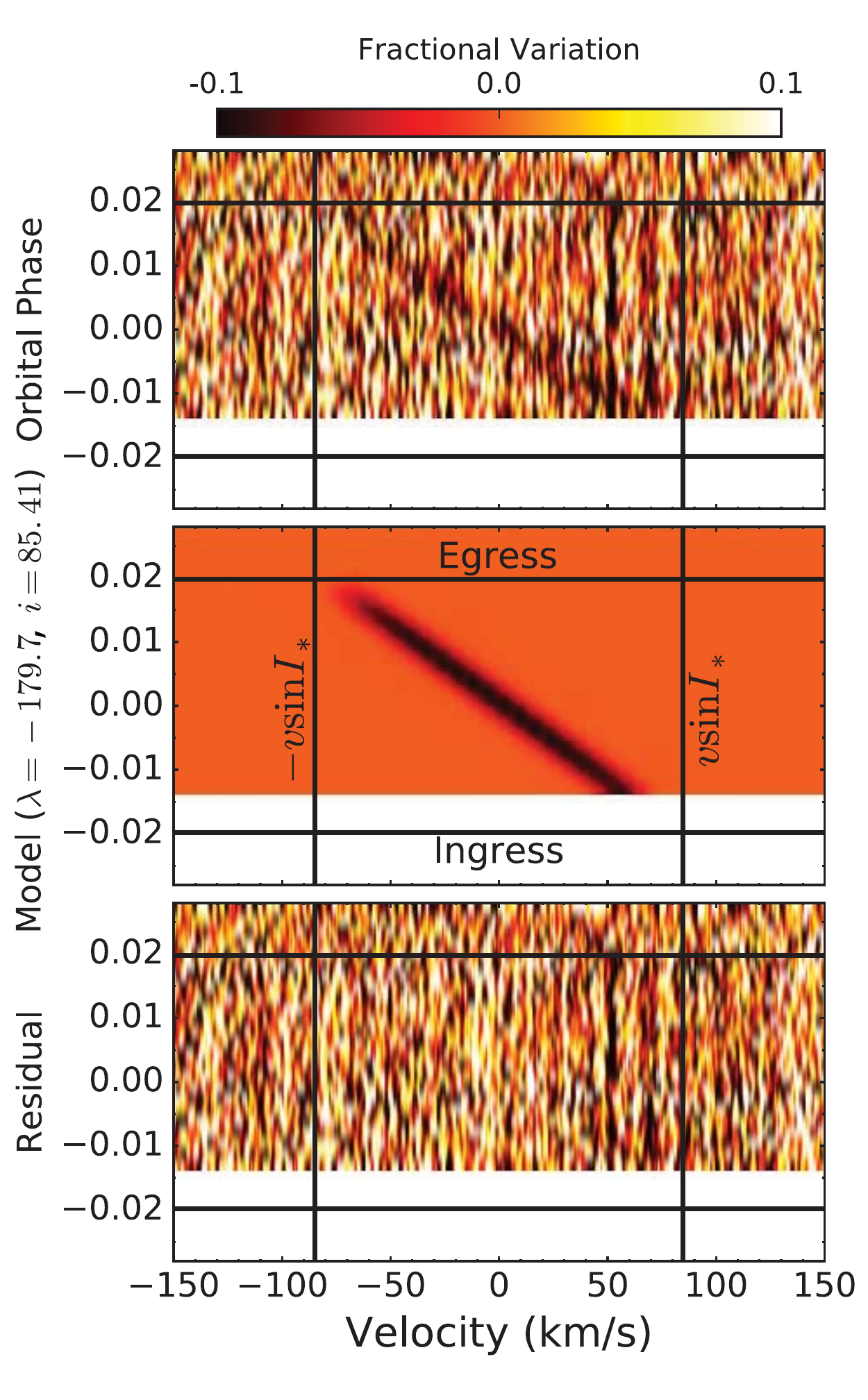}
\caption{Doppler tomographic line profile plot. The top panel shows the spectroscopic data, the middle panel shows the derived model, and the bottom panel shows the residuals. In each panel, the vertical lines denote the width of the convolution kernel (i.e., $\pm\vsini$), and the horizontal lines show the duration of the transit. Time increases from bottom to top. Each color-scale row indicates the deviation of the line profile at that time from the out-of-transit line profile, with dark regions of the plot indicating regions of the in-transit line profile that are shallower with respect to the out-of-transit line profile. The Doppler tomographic signal implies a retrograde orbit for the planet, as the line profile perturbation moves from the red wing of the line profile across to the blue wing. The planet moves in a corresponding manner during the transit, from obscuring the red-shifted hemisphere of the star across to the blue-shifted hemisphere. 
\label{fig:dt}}
\end{center}
\end{figure}

To confirm that a transiting companion is indeed orbiting the primary star in the KELT-19 system, and to measure the projected spin-orbit angle and impact parameter of the planetary orbit, we performed a Doppler tomographic analysis of the spectroscopic transit observed by TRES on 2016-11-08. Line broadening profiles were derived for each observation via a least-squares deconvolution analysis \citep[following][]{Donati1997,CollierCameron2010,Zhou:2016}. A master broadening profile was calculated by combining the out-of-transit profiles. Each observation was then subtracted from the master broadening profile, revealing the spectroscopic shadow of the transiting planet, as shown in Figure~\ref{fig:dt}. The Doppler tomographic signal was modelled as per \citet{Gaudi:2017}. Limb darkening parameters were adopted from \citet{Claret:2004} for the photometric $V$ band, similar to the wavelength region from which the broadening profiles were derived. 

\subsubsection{Stellar Parameters from Spectra}\label{sec:stellarPars}

Because the spectrum of KELT-19A includes the light
from KELT-19B (see \S \ref{sec:absRV} and \S \ref{sec:ao}),
standard spectral synthesis or fitting techniques that ignore the
influence of the secondary on the primary line profiles may be
susceptible to systematic bias. We therefore applied a two-dimensional
cross-correlation analysis \citep[TODCOR;][]{Zucker:1994} using pairs of synthetic spectra to
identify the stellar parameters that provided the best fit to the
observed composite spectra. For this analysis we used the TRES spectra
of KELT-19 and the CfA library of synthetic spectra, which were
generated by John Laird using Kurucz model
atmospheres \citep{Kurucz:1992} and a linelist compiled by Jon
Morse. The synthetic grid covers the wavelength range
$5050$--$5350$\ {\AA}, and has spacing of $250$\ K in \teff\ and $0.5$\
dex spacing in $\loggstar$\ and \mh. We note that this latter parameter is
a scaled solar bulk metallicity, rather than the iron
abundance, \feh. It is generally a reasonable assumption that the two
quantities are similar, but it might not be the case for stars
exhibiting peculiar abundances (like many A stars). Throughout the
paper we do use \mh\ and \feh\ interchangeably, but because we neither
derive nor impose strong constraints on the metallicity, we expect
that any differences between the two quantities will have negligible effects on
our results.

We ran TODCOR on all combinations of templates in the (6-dimensional)
parameter space spanning temperatures $6000 \le T_{\rm eff,A} \le          
8500$\,K and $3750 \le T_{\rm eff,B} \le 6750$\,K, surface
gravities $3.0 \le \loggstar \le 5.0$, and metallicities $-1.5\ {\rm  
dex} \le \mh \le +0.5$\ dex for both stars. We allowed the primary
and secondary metallicities to be fit independently because even if
the two stars formed together, many A stars display peculiar
photospheric metallicities. The mean TODCOR correlation coefficient
from each of these $\mysim 37000$\ template pairs defines a $6$-D
surface (the axes corresponding to the $6$\ stellar parameters), on
which we interpolate to the peak and adopt the corresponding stellar
parameters. The result comes with several caveats. Derived
spectroscopic stellar parameters are highly covariant ---
temperatures, metallicities, and gravities can be altered
simultaneously to obtain very similar spectra over relatively large
ranges of parameter space --- so this degeneracy must be broken with
independent constraints. In our case, we have derived the primary
surface gravity ($\loggstar = 4.127$) from constraints on the stellar density, mass, and radius as part of the global system fit (see \S \ref{sec:globalfit}). Because $\loggstar$\ is determined so precisely, even a $3$-$\sigma$\ error in this value has minimal effect on the other parameters. As a result, we fix it in the TODCOR analysis for simplicity. Additionally, the secondary spectrum
possesses a very low signal-to-noise ratio, so its parameters are
poorly constrained by the spectra alone. Instead, we require it to be
a main sequence companion $\loggstar \mysim 4.5$, with a temperature of
$5200$\ K (as derived in our initial SED
analysis; \S \ref{sec:SED}). We note that the projected rotational
velocities, \vsini, are nearly orthogonal to the other parameters, so
we fixed these to simplify the analysis and reduce computation time:
the primary \vsini\ was set to $84.1$\kms\ (see \S \ref{sec:absRV}),
while the secondary \vsini\ was estimated to be $\mysim 2$\kms\ via
an empirical gyrochronology relation \citep{Mamajek:2008} and the age
and colors derived from the initial SED and isochrone analysis.

Under these constraints, we find the following parameters: $T_{\rm
eff,A} = 7505 \pm 104$\ K; ${\rm [m/H]_{A}} = +0.24 \pm 0.16$; ${\rm
[m/H]_{B}} = -0.26 \pm 0.35$. The reported errors include
contributions from both formal and correlated errors. It is
interesting to note that the primary metallicity is $0.5$\ dex higher
than that of the secondary, albeit at low confidence because of the
noisy secondary spectrum. We would expect to observe this difference
if the primary is an Am star, a possibility we explore in
\S \ref{sec:Am}. Given the uncertainty in the metallicities ---
and the possibility that the photospheric spectrum of the primary
is not representative of its true metallicity --- we choose to adopt a
broad metallicity prior appropriate for the solar neighborhood ($\feh =
0.0 \pm 0.5$\ dex) in our subsequent global modeling. The main result
of the TODCOR analysis, then, is a spectroscopic temperature for the
primary of $\teff=7505 \pm 104$\ K.

\subsubsection{KELT-19A is likely an Am star}\label{sec:Am}

As noted in the introduction, KELT-19A has a peculiar abundance pattern that is indicative of it belonging to the class of metallic-line A stars (Am stars).  The hallmark of such stars is that they have some stronger metallic lines (such as strontium) than are expected for stars of their effective temperatures (as measured by their, e.g., H$\alpha$ line), but weaker lines in others, such as calcium, than expected for the same metallicity and effective temperature.  In other words, the star does not appear to have a consistent metallicity given its effective temperature.

\begin{figure*}[!ht]
\includegraphics[width=1.0\linewidth]{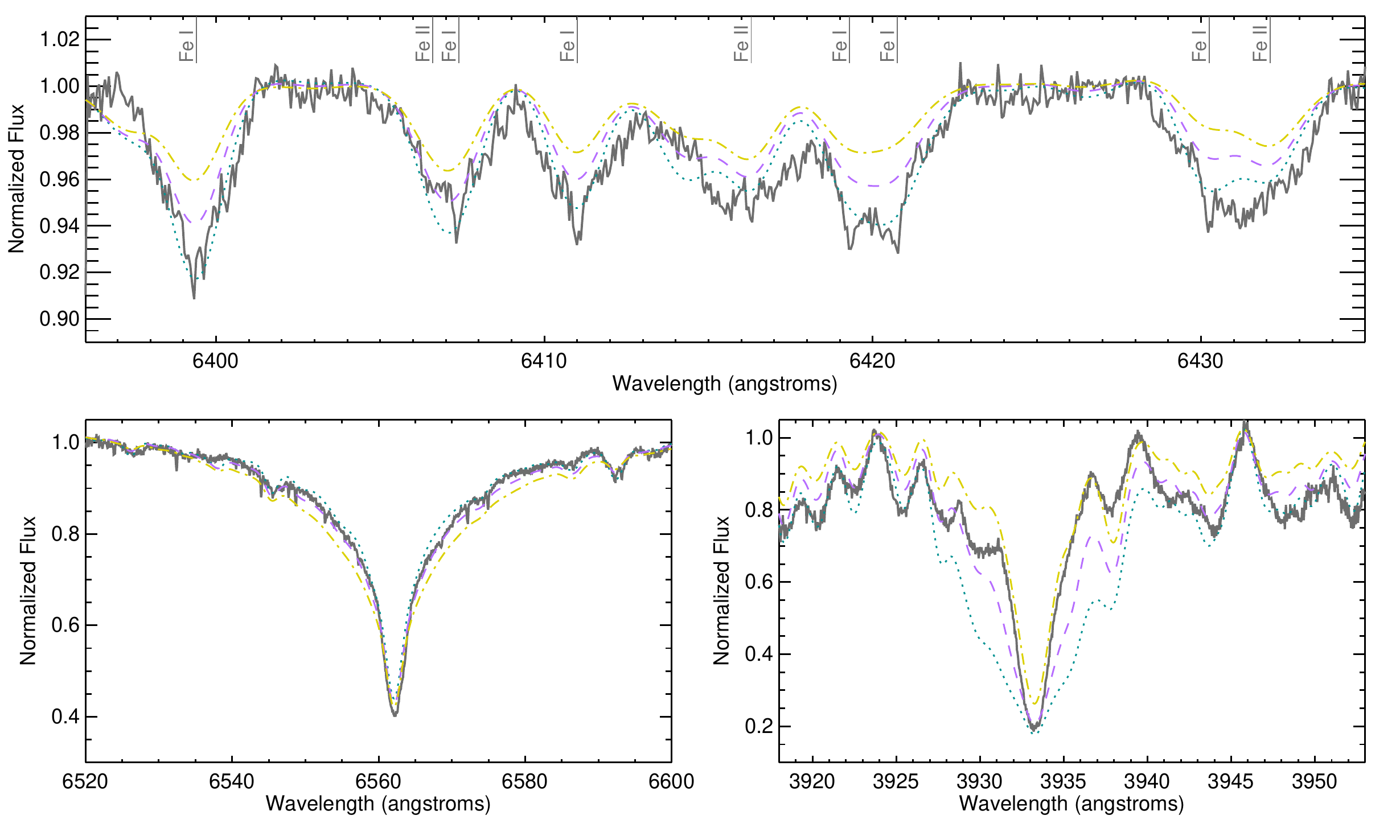}
\caption{ The spectrum of KELT-19A (gray solid line) and three PHOENIX model atmospheres \citep[colored lines;][]{Husser:2013}, overplotted for illustration (i.e., not fit to the data). All models were generated with $\loggstar = 4.124$\ (cgs) and ${\rm [Fe/H]} = +0.5$ and were broadened to $84.8$\kms\ rotation, but have temperatures of $7000$\ K (blue-green dotted), $7500$\ K (purple dashed), and $8000$\ K (yellow dash-dotted). {\it Bottom left:} The H$\alpha$ profile is consistent with a $7500$\ K atmosphere, like we find in the spectroscopy and the global fit. {\it Top:} Iron lines are enhanced, and therefore more consistent with a cooler ($7000$\ K) atmosphere. {\it Bottom right:} The \ion{Ca}{2} K line is weaker than expected, with a profile similar to that of the $8000$\ K atmosphere. A spectral type that is ``too late'' in metals and a \ion{Ca}{2} K spectral type that is ``too early'' for the Balmer line spectral type is a hallmark of Am stars because of their photospheric metal enhancement and calcium deficiency.}
\label{fig:Am}
\end{figure*}

\begin{figure*}[!ht]
\includegraphics[width=1.0\linewidth,trim=5 -3 10 0,clip=true]{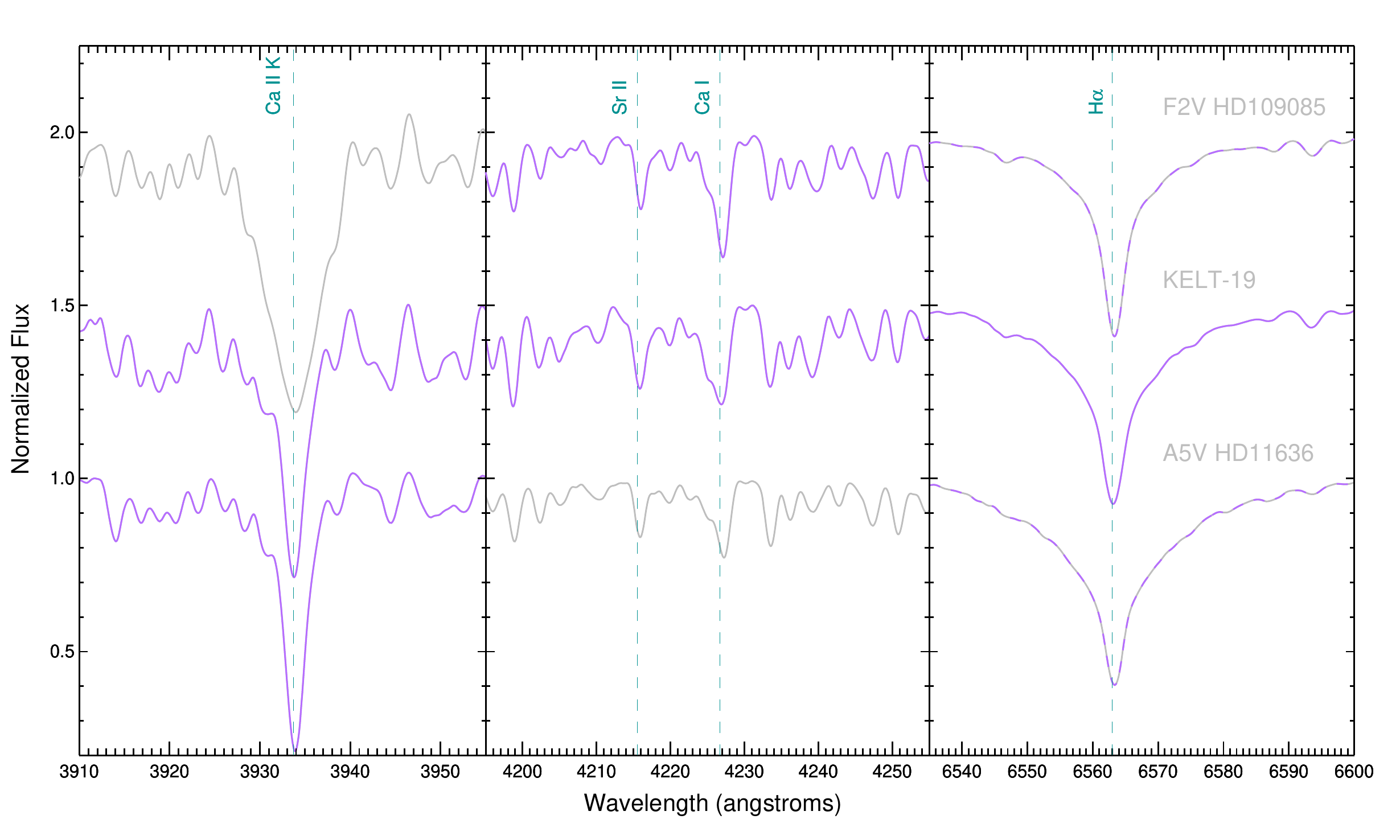}
\caption{The spectrum of KELT-19 (middle row, purple) and two spectral standard stars -- F2V HD109085 (top) and A5V HD11636 (bottom). All spectra were obtained by TRES and broadened so that the total broadening (combined rotational, instrumental and artificial) is 100\kms. Each panel shows a feature or features important for determining spectral type, and we color the matching spectrum purple to indicate a match (or dashed purple to indicate a partial match). The \ion{Ca}{2} K profile (left) of KELT-19 is an excellent match for that of the A5V star. The H$\alpha$ profile (right) is broader than the F2V, narrower than the A5V, and matches well to an A7V spectrum (not pictured). The strength of the metal lines (middle) match the F2V spectrum, with the exception of \ion{Ca}{1}, which is much weaker in KELT-19. \ion{Sr}{2}, which is expected to be enhanced in Am stars, does not appear significantly stronger in KELT-19, but it is blended with many other lines because of the star's rapid rotation.}
\label{fig:SpT}
\end{figure*}

This leads to a classical definition of Am stars, which notes that the spectral type one deduces depends on the feature used for typing. Because A stars in general have metallic lines that increase in strength toward later type, a spectral type based on some metal lines that show enhancement will lead to a spectral type for an Am star that is ``too late'' compared to the Balmer line spectral type.  Similarly, because A stars have Ca II K lines that increase in strength toward later type, the calcium deficiency for Am stars will lead to a calcium spectral type that is ``too early''.  Thus a classical definition of Am stars is a range of spectral types from these methods of at least 5 subtypes. This is demonstrated in the spectrum of KELT-19A in Figures \ref{fig:Am} and \ref{fig:SpT}. 

In Figure \ref{fig:Am}, the top panel shows iron lines, bottom left shows H$\alpha$, and bottom right shows the Ca II K line. In each panel, the black line is the observed KELT-19A spectrum, and the thin colored lines are three PHOENIX model atmospheres. Each has KELT-19A's estimated $\loggstar$, $\vsini$ and [Fe/H]=+0.5. The blue-green dotted, purple dashed, and yellow dash-dotted lines correspond to 7000, 7500, and 8000 K, respectively\footnote{None of these are fits; they are merely overplotted for illustration.}.  One can see that the 7000 K model (blue-green dotted line) is most appropriate for the metal lines, whereas the H$\alpha$ is most consistent with our adopted temperature ($\sim$7500 K; purple dashed line), and the Ca II K line profile is most consistent with a hotter star (8000 K; yellow dash-dotted line). We also note that solar metallicity PHOENIX models all yield \ion{Fe}{2} lines that are too weak at any temperature, which provides additional evidence that the photospheric metallicity is enhanced, as hinted at in the TODCOR analysis of \S \ref{sec:stellarPars}. 

To provide a more detailed spectral type for KELT-19, we compare its spectrum to a sequence of observed spectra ranging from A3V to F3V. All of these spectra were observed by TRES and reduced in the same way as KELT-19, which minimizes systematic bias, e.g. due to continuum normalization. Because each star has a different projected rotational line broadening, we measure it for each star and convolve the normalized spectrum with a Gaussian with a width appropriate to produce a total broadening (rotational, instrumental, and artificial) of $100$\kms. We compare KELT-19 to the spectral sequence and identify the spectral types that provide the best match to the \ion{Ca}{2} K, H$\alpha$, and metal lines of KELT-19, and we illustrate this in Figure \ref{fig:SpT}. While the \ion{Sr}{2} line does not show any obvious enhancement, as might be expected for an Am star, this is not entirely surprising: the abundance anomalies in Am stars are negatively correlated with rotation so that those rotating as rapidly as KELT-19 are less anomalous; and the relatively rapid rotation of KELT-19 results in the \ion{Sr}{2} line blending with at least three other lines of similar strength, so that the strength of any anomaly is diluted. Nevertheless, the other features in the spectrum are consistent with an Am star. An A5V star is an excellent match for the \ion{Ca}{2} K line, an A7V star for the H$\alpha$\ profile, and an F2V star is the best match for the strength of the metal lines, resulting in a range of spectral types of $\sim$A5 to $\sim$F2. We therefore conclude that KELT-19A meets the classical definition of an Am star, with a spectral type of ``Am kA5 hA7 mF2 V''.

\begin{table}
\begin{center}
\caption{Radial Velocity Measurements of KELT-19\label{tab:rv1}}
\begin{tabular}{lcccc}
\tableline
\multicolumn{1}{c}{\bjdtdb} & \multicolumn{1}{c}{Primary} & \multicolumn{1}{c}{Primary} & \multicolumn{1}{c}{Secondary} & \multicolumn{1}{c}{Secondary}\\
\multicolumn{1}{c}{} & \multicolumn{1}{c}{RV} & \multicolumn{1}{c}{$\sigma_{RV}$\tablenotemark{a} } & \multicolumn{1}{c}{RV} & \multicolumn{1}{c}{$\sigma_{RV}$\tablenotemark{a}}\\
\multicolumn{1}{c}{} & \multicolumn{1}{c}{(m\,s$^{-1}$)} & \multicolumn{1}{c}{(m\,s$^{-1}$)} & \multicolumn{1}{c}{(m\,s$^{-1}$)} & \multicolumn{1}{c}{(m\,s$^{-1}$)}\\
\tableline
\multicolumn{5}{l}{TRES} \\
\vspace{-0.1in}\\
~~~2457118.717801 &  -8322 &  677 &  -6315 & 2185 \\
~~~2457323.926960 &  -7382 &  817 &  -9178 & 1588 \\
~~~2457704.974522 &  -7582 &  378 &  -9499 & 1054 \\
~~~2457706.006779 &  -8353 &  379 &  -9178 &  928 \\
~~~2457706.905449 &  -8185 &  443 &  -9129 &  951 \\
~~~2457715.859122 &  -7872 &  422 &  -9260 &  971 \\
~~~2457761.845292 &  -8667 &  319 &  -8922 &  871 \\
\multicolumn{5}{l}{McDonald} \\
~~~2457685.865924 &  -7126 &  413 &  -9227 &  509 \\
~~~2457687.904483 &  -6797 &  434 &  -8635 &  391 \\
~~~2457732.803281 &  -7243 &  266 &  -9069 &  330 \\
~~~2457732.925982 &  -7461 &  262 &  -8580 &  271 \\
~~~2457733.004039 &  -7236 &  273 &  -8784 &  251 \\
~~~2457733.890461 &  -7329 &  552 &  -8256 &  862 \\
~~~2457734.795794 &  -7234 &  304 &  -9021 &  312 \\
~~~2457734.998676 &  -7294 &  253 &  -8503 &  241 \\
~~~2457735.810347 &  -7199 &  251 &  -8656 &  298 \\
~~~2457736.002794 &  -7517 &  366 &  -8834 &  326 \\
~~~2457736.816475 &  -7434 &  244 &  -8366 &  342 \\
~~~2457737.008672 &  -6577 &  297 &  -8067 &  395 \\
~~~2457737.771236 &  -7800 &  263 &  -9017 &  268 \\
~~~2457738.009789 &  -7263 &  331 &  -8445 &  322 \\
\tableline
\end{tabular}
\tablecomments{Because of the rapidly rotating host star, we were unable to derive bisector spans.}
\vspace{-10 pt}
\tablenotetext{1}{RV errors before being scaled by \multifast.}
\end{center}
\label{tbl:RVs}
\end{table}

\section{Host Star Properties}\label{sec:star_props}

Table \ref{tbl:LitProps} lists various properties and measurements of KELT-19 collected from the literature and derived in this work. The data from the literature include  $BV$ and $gri$ photometry from \citet{Henden:2015}, optical fluxes in the $B_{\rm T}$ and $V_{\rm T}$ passbands from the Tycho-2 catalog \citep{Hog:2000}, near-infrared (IR) fluxes in the $J$, $H$ and $K_{\rm S}$ passbands from the 2MASS Point Source Catalog \citep{Cutri:2003,Skrutskie:2006}, near- and mid-IR fluxes in four WISE passbands \citep{Wright:2010,Cutri:2012}, and distance and proper motions from Gaia \citep{Brown:2016}.

\begin{table}
\footnotesize
\centering
\caption{Literature Properties for KELT-19}
\begin{tabular}{llcc}
  \hline
  \hline
Other Names\dotfill & 
        \multicolumn{3}{l}{BD+07 1721} \\
      & \multicolumn{3}{l}{TYC 764-1494-1}				\\
	  & \multicolumn{3}{l}{2MASS J07260228+0736569} \\
	  & \multicolumn{3}{l}{TIC 425206121}
\\
\hline
Parameter & Description & Value & Ref. \\
\hline
$\alpha_{\rm J2000}$\dotfill	&Right Ascension (RA)\dotfill & $07^h26^m02\fs2895$			& 1	\\
$\delta_{\rm J2000}$\dotfill	&Declination (Dec)\dotfill & +07\arcdeg36\arcmin56\farcs834			& 1	\\
\\
$B_{\rm T}$\dotfill			&Tycho $B_{\rm T}$ mag.\dotfill & $10.273 \pm 0.036$		& 2	\\
$V_{\rm T}$\dotfill			&Tycho $V_{\rm T}$ mag.\dotfill & $9.899 \pm 0.035$		& 2	\\

$B$\dotfill		& APASS Johnson $B$ mag.\dotfill	& $10.201 \pm 0.030$		& 3	\\
$V$\dotfill		& APASS Johnson $V$ mag.\dotfill	& $9.885  \pm 0.040$		& 3	\\

$g'$\dotfill		& APASS Sloan $g'$ mag.\dotfill	&  $10.163 \pm 0.120$		& 3	\\
$r'$\dotfill		& APASS Sloan $r'$ mag.\dotfill	& 	$9.872 \pm 0.050$       & 3	\\
$i'$\dotfill		& APASS Sloan $i'$ mag.\dotfill	& 	$9.878 \pm 0.040$	& 3	\\
\\
$J$\dotfill			& 2MASS $J$ mag.\dotfill & $9.343 \pm 0.030$		& 4	\\
$H$\dotfill			& 2MASS $H$ mag.\dotfill & $9.237 \pm 0.020$	& 4	\\
$K$\dotfill			& 2MASS $K$ mag.\dotfill & $9.196 \pm 0.020$	& 4	\\
\\
\textit{WISE1}\dotfill		& \textit{WISE1} mag.\dotfill & $9.138 \pm 0.022$		& 5	\\
\textit{WISE2}\dotfill		& \textit{WISE2} mag.\dotfill & $9.156 \pm 0.020$		& 5 \\
\textit{WISE3}\dotfill		& \textit{WISE3} mag.\dotfill & $9.132 \pm 0.035$		& 5	\\
\textit{WISE4}\dotfill		& \textit{WISE4} mag.\dotfill & $\ge8.089$ 		& 5	\\
\\
$\mu_{\alpha}$\dotfill		& Gaia DR1 proper motion\dotfill & -3.706 $\pm$ 1.126 		& 6 \\
                    & \hspace{3pt} in RA (mas yr$^{-1}$)	& & \\
$\mu_{\delta}$\dotfill		& Gaia DR1 proper motion\dotfill 	&  -1.303 $\pm$ 1.226 &  6 \\
                    & \hspace{3pt} in DEC (mas yr$^{-1}$) & & \\
\\
$RV$\dotfill & Systemic radial \hspace{9pt}\dotfill  & $-8.5\pm0.5$ & \S\ref{sec:absRV} \\
     & \hspace{3pt} velocity (\kms)  & & \\
$v\sin{i_\star}$\dotfill &  Stellar rotational \hspace{7pt}\dotfill &  $84.8\pm2.0$ & \S\ref{sec:globalresults} \\
                 & \hspace{3pt} velocity (\kms)  & & \\
Sp. Type$_A$\dotfill & Primary Star Sp. Type\dotfill & A8V & \S\ref{sec:stellarPars} \\ 
Sp. Type$_B$\dotfill & Secondary Star Sp. Type\dotfill & G9V--K1V & \S\ref{sec:stellarPars} \\
Age\dotfill & Age (Gyr)\dotfill & $1.1\pm0.1$ & \S\ref{sec:hrd_and_age} \\
$\Pi$\dotfill & Gaia Parallax (mas) \dotfill & 3.60 $\pm$ 0.72 & 6\dag \\ 
$d_{\star Gaia}$\dotfill& Gaia-inferred dist. (pc) \dotfill & $278^{+69}_{-47} $ & 6\dag \\
$d_{\star SED}$\dotfill& SED-inferred dist. (pc) \dotfill & $255 \pm 15$ & \S\ref{sec:SED} \\
$A_V$\dotfill & Visual extinction (mag) & $0.03 \pm 0.03$ & \S\ref{sec:SED} \\
$U^{*}$\dotfill   & Space motion (\kms)\dotfill & $14.6 \pm 0.9$ & \S\ref{sec:UVW} \\
$V$\dotfill       & Space motion (\kms)\dotfill & $17.6 \pm 1.3$ & \S\ref{sec:UVW} \\
$W$\dotfill       & Space motion (\kms)\dotfill & $ 0.2 \pm 1.4$ & \S\ref{sec:UVW} \\
\hline
\hline
\end{tabular}
\begin{flushleft} 
 \footnotesize{ \textsc{NOTES:}
    References are: $^1$\citet{vanLeeuwen:2007},$^2$\citet{Hog:2000}, $^3$\citet{Henden:2015}, $^4$\citet{Cutri:2003}, $^5$\citet{Cutri:2013},$^6$\citet{Brown:2016} Gaia DR1 http://gea.esac.esa.int/archive/
    \dag Gaia parallax after correcting for the systematic offset of $-0.18$~mas for an ecliptic latitude of $-14\degr$ as described in \citet{Stassun:2016}.
}
\end{flushleft}
\label{tbl:LitProps}
\end{table}

\section{Analysis and Results}

\subsection{SED Analysis}\label{sec:SED}
\begin{figure}[!ht]
\includegraphics[width=\linewidth,trim=95 50 90 80,clip=true]{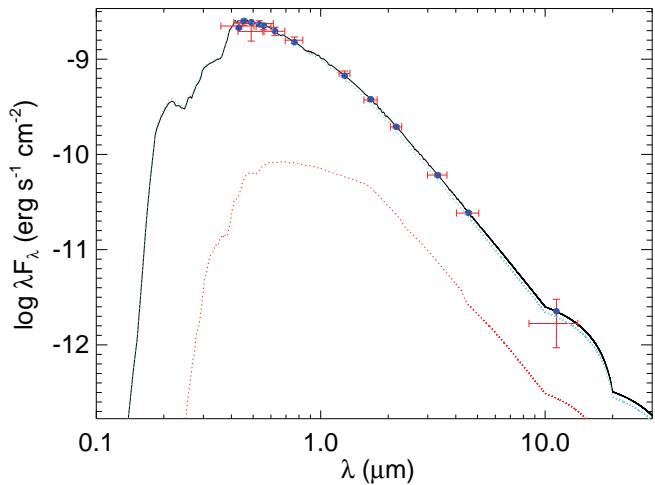}
\caption{KELT-19 two-component spectral energy distribution (SED). Crosses represent the measured fluxes of the two unresolved stars, with vertical bars representing the measurement uncertainties and the horizontal bars representing the width of the bandpass. The blue dots are the predicted passband-integrated fluxes of the best-fit theoretical SED corresponding to our observed photometric bands. The black solid, blue dotted, and red dotted curves represent the best-fit two-component, KELT-19A, and KELT-19B stellar atmospheres, respectively, from \citet{Kurucz:1992} (see the text).}
\label{fig:sed}
\end{figure}

We performed a fit to the broadband spectral energy distribution (SED) of KELT-19 in order to obtain constraints on stellar parameters for use in the global system fit. We assembled the available broadband photometry from extant catalogs, with measurements spanning over the wavelength range 0.4--22~\micron\ (see Figure~\ref{fig:sed} and Table~\ref{tbl:LitProps}). 

For the fitting, we used the stellar atmosphere models of \citet{Kurucz:1992}, where the free parameters are the effective temperature (\teff), the extinction (\av), and the distance (\dstar). In principle, the atmosphere models also depend on metallicity (\feh) and surface gravity (\loggstar), however we do not have strong independent constraints on these, and in any event they are of secondary importance to \teff\ and \av. Thus we assumed a main-sequence \loggstar$\approx$4.0 and a solar \feh. For \av, we restricted the maximum permitted value to be that of the full line-of-sight extinction from the dust maps of \citet{Schlegel:1998}. We also ran the fit with \feh = +0.5 and the result was not significantly different than the solar metallicity result.

Importantly, the high-resolution imaging (see \S \ref{sec:ao}) revealed another faint star, sufficiently close to KELT-19 that it can be assumed to contaminate the broadband photometry. Therefore, we performed the fit with two components, assuming (for the purposes of the fit) the same \av\ and \dstar\ for both, and we adopted as additional constraints the flux ratios determined from the adaptive optics imaging and from the spectroscopic analysis: $F_{\mathrm{B}}/F_{\mathrm{A}} = 0.0270 \pm 0.0034$ in the range $5200\pm150$\,\AA, $\Delta J = 2.50 \pm 0.06$, and $\Delta K_S = 2.045 \pm 0.030$. This introduces one additional fit parameter, namely, the ratio of stellar radii ($R_{\mathrm B} / R_{\mathrm A}$) that effectively sets the relative bolometric fluxes of the two stars. 

The best fit model shown in Figure~\ref{fig:sed} has a reduced $\chi^{2}$ of 0.66. We find $A_{\rm V}$ = $0.03 \pm 0.03$, ${\teff}_{\mathrm{A}} = 7500 \pm 200$~K, ${\teff}_{\mathrm{B}} = 5200 \pm 100$~K, $\dstar = 255 \pm 15$~pc, and $R_2 / R_1 = 0.46 \pm 0.03$.
We note that the quoted statistical uncertainties on $A_{\rm V}$ and $\teff$ are
likely to be slightly underestimated because we have not accounted for the
uncertainty in $\loggstar$ or \feh. We also note, however, that the inferred \dstar\ obtained here is fully consistent with that from the {\it Gaia\/} parallax \citep[after correction for the systematic offset of $-0.18$ mas determined by][]{Stassun:2016}, and moreover the inferred properties of the secondary star are consistent with those of the observed secondary spectrum (see \S \ref{sec:spectroscopy}). 

The two-component SED fit also permits determination of the amount of contaminating flux from the companion in the observed transit at each wavelength. This is accounted for in the global solution as discussed in \S \ref{sec:globalfit}.

\subsection{Stellar Models and Age}\label{sec:hrd_and_age}

\begin{figure}[!ht]
\includegraphics[width=2.7in,angle=90,trim=10 -5 25 50,clip]{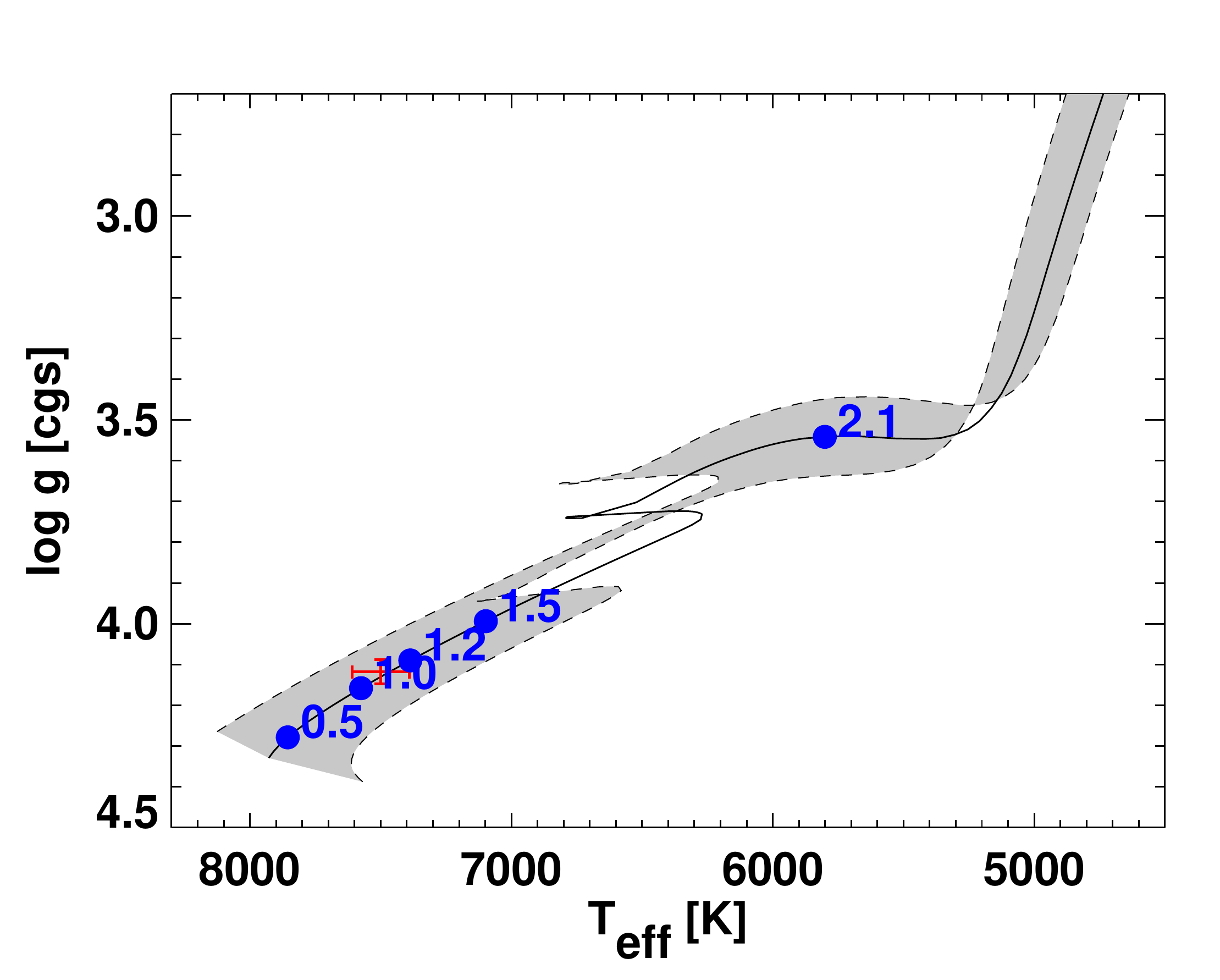}
\caption{Evolution of the KELT-19A system in the Kiel diagram. The red cross represents the KELT-19A parameters from the final global fit. The black curve represents the theoretical evolutionary track for a star with the mass and metallicity of KELT-19A, and the grey swath represents the uncertainty on that track based on the uncertainties in mass and metallicity. Nominal ages in Gyr are shown as blue dots.}
\label{fig:hrd}
\end{figure}

With $\teff$ from the SED analysis, and with an estimated \loggstar\ and \Mstar\ from the global analysis (see below), we can place KELT-19A in the Kiel diagram for comparison with theoretical stellar evolutionary models (Fig.~\ref{fig:hrd}). 
The estimated system age using the final global fit parameters is $\approx$1.1~Gyr, with an approximate uncertainty of order 0.1~Gyr. The KELT-19 system is more than halfway through its main-sequence lifetime but is at a stage of evolution well before the ``blue hook'' transition to the subgiant and eventual red giant evolutionary phases. \\[0.2in]

\subsection{UVW Space Motion}\label{sec:UVW}

We determine the three-dimensional space velocity of KELT-19 in the usual $(U,V,W)$ coordinates in order to determine the Galactic population to which it belongs. We used a modification of the IDL
routine {\tt GAL\_UVW}, which is itself based on the method of \citet{Johnson:1987}. We adopt the Gaia proper motions as listed in Table \ref{tbl:LitProps}, the SED-inferred distance $255 \pm 15$\,pc, and the absolute radial velocity as determined from TRES spectroscopy of $-8.5\pm0.5$\kms. We find that KELT-19 has $U,V,W$ space motion of $(U,V,W)= (14.6 \pm 0.9, 17.6 \pm 1.3, 0.2 \pm 1.4)$\kms, in a coordinate system where positive $U$ is in the direction of the Galactic center, and using the \citet{Coskunoglu:2011} determination of the solar motion with respect to the local standard of rest.  These values yield a 99.2\% probability that the KELT-19 binary system is in the thin disk, according to the classification scheme of \citet{Bensby:2003}, as expected for its age and spectral type.

\subsection{Global System Fit}\label{sec:globalfit}

We determined the physical and orbital parameters of the KELT-19A system by jointly fitting 8 transit light curves, 7 TRES and 14 HJST out-of-transit RVs, and a TRES Doppler tomographic data set (see \S \ref{sec:spectroscopy}). To perform the global fit, we used {\tt MULTI-EXOFAST} (\multifast\ hereafter), which is a custom version of the public software package {\tt EXOFAST} \citep{Eastman:2013}. \multifast\ first performs an {\tt AMOEBA} \citep{Nelder:1965} best fit to each of the RV and light curve data sets individually to determine uncertainty scaling factors. The uncertainties are scaled such that the probability that the \chisq \ for a data set is larger than the value we achieved, $\pchisq$, is $0.5$, to ensure the resulting parameter uncertainties are roughly accurate. The resulting RV uncertainty scaling factors are 1.22 and 1.13 for the TRES and HJST velocities, respectively. The uncertainties of the DT observations were scaled by 1.0. Finally, \multifast\ performs a joint {\tt AMOEBA} model fit to all of the datasets and executes a Markov Chain Monte Carlo (MCMC), starting at the global best fit values, to determine the median and 68\% confidence intervals for each of the physical and orbital parameters.  \multifast\ provides the option to include the Yonsei-Yale (YY) stellar model constraints \citep{Demarque:2004} or the Torres empirical constraints \citep{Torres:2010} to break the well-known degeneracy between \Mstar\ and \Rstar\ for single-lined spectroscopic eclipsing systems. \citet{Siverd:2012} provides a more detailed description of \multifast, except the Doppler tomographic model is described in \citet{Gaudi:2017}.

\subsubsection{Light Curve Detrending and Deblending}

 We use {\tt AIJ} to determine the best detrending parameter data sets to include in the \multifast\ global model by finding the {\tt AMOEBA} best fit of a \citet{Mandel:2002} exoplanet transit model to the transit photometry plus linear combination(s) of detrending data set(s). Up to two detrending data sets were selected per light curve based on the largest reductions in the Bayesian Information Criterion (BIC) calculated by {\tt AIJ} from the model fits with and without the detrending data set included. A detrending data set was not included unless it reduced the BIC by $>2.0$, resulting in some light curves with only one detrending data set. The final detrending data sets we chose for each light curve are listed in Table \ref{tbl:photobs}. It is important to emphasize that the {\tt AIJ}-extracted raw differential light curves (i.e. not detrended) and the detrending data sets were inputs to \multifast\ and were simultaneously fitted as a part of the global models.
 
 As discussed in \S \ref{sec:ao} and \S \ref{sec:absRV}, KELT-19A has a bound stellar secondary companion at a projected separation of $0\farcs64$. Because the secondary is blended in all follow-up photometry apertures, the flux from the secondary must be taken into account to obtain the correct transit depth and planetary radius (e.g. \citealt{Ciardi:2015}). As discussed in \S \ref{sec:SED}, the two-component SED fit permits determination of the amount of contaminating flux from the companion in the observed transit at each wavelength. The determined blend factors, $F2/F1$, for all of the follow-up photometry filter bands are shown in Table \ref{tbl:fluxContamination}. The blend factors for each filter band were included in \multifast\ as fixed values to adjust the transit depth in each filter to account for the blend.   

\begin{table}
 \centering
 \caption{Flux Contamination From SED Fit}
 \label{tbl:fluxContamination}
 \begin{tabular}{cccc}
    \hline
    \hline
    \multicolumn{1}{c}{Band} & \multicolumn{1}{c}{$F_{\textrm B}/F_{\textrm A}$} \\
    \hline
    $5200\pm150$\,\AA & 0.02782\\
    U           &   0.02412 \\
    B           &   0.03389 \\
    V           &   0.03710 \\
    R           &   0.04619 \\
    I           &   0.04693 \\
    Sloan $g'$  &   0.03961 \\
    Sloan $r'$  &   0.04469 \\
    Sloan $i'$  &   0.04867 \\
    Sloan $z'$  &   0.05261 \\
    \hline
    \hline
 \end{tabular}
\end{table}

\subsubsection{Gaussian and Uniform Priors}\label{sec:priors}

We included Gaussian priors on the reference transit center time, $T_0$, and orbital period, $P$. To determine the prior values for the final global fits, we executed preliminary \multifast\ global fits, including a TTV parameter in the model for each light curve to allow the transit center time to vary from a linear ephemeris, and used priors $T_0 = 2457055.276\pm0.013$ \bjdtdb\ and $P=4.611758\pm0.000053$~d derived from the KELT data. For these preliminary fits, we included the 8 primary transit light curves and the DT data. We chose to include a circular orbit constraint and fixed the RV slope to zero for the model fits. The preliminary YY-constrained model fit resulted in a TTV-based linear ephemeris $T_0 =2457281.249522\pm0.000359$ \bjdtdb\ and $P=4.6117091\pm0.0000089$\,d. These values were used as Gaussian priors in the final YY-constrained global model fit. The preliminary Torres-constrained model fit resulted in a TTV-based linear ephemeris $T_0 =2457285.861243\pm0.000355$ \bjdtdb\ and $P=4.6117094\pm0.0000090$\,d. These values were used as Gaussian priors in the final Torres-based global model fit. Since the KELT- and TTV-based ephemerides are generally derived from independent data, we propagate forward the precise TTV-based ephemerides without concern for double-counting data.

We also included Gaussian priors on the stellar parameters $\teff=7505\pm104$\,K and $\feh=0.0\pm0.5$ from the SED analysis in \S \ref{sec:SED} and the stellar parameter analysis in \S \ref{sec:stellarPars} and $v\sin I_\star = 84.1 \pm 2.1$\kms\ and macroturbulent broadening of $3.4 \pm 2.0$\kms\ from the out-of-transit broadening profile. A prior was not imposed on \loggstar, since the value derived from the light curve-based stellar density and our stellar radius constraints is expected to be more accurate than the spectroscopic (e.g. \citealt{Mortier:2013,Mortier:2014}) or SED-based \loggstar.

We limited the range of certain parameters by including bounded uniform priors. We restricted the RV semi-amplitude to $K > 1.0\,\mathrm{m\,s}^{-1}$. To prevent problems when interpolating values from the limb darkening tables, we restricted the stellar parameters to $3500 \le \teff < 20{,}000$ K, $-2.0 \le \feh < 1.0$, and $2.0 \le \loggstar <5.0$. We inspected the corresponding posterior parameter distributions to ensure that there was no significant likelihood near the uniform prior boundaries.

\subsubsection{Global Model Configurations}\label{sec:models}

We examine the results of two global model configurations to explore the effects of YY-constrained and Torres-constrained global model fits on parameter posterior distributions. Since no RV orbit is detected, we force both models to have a circular orbit and an RV slope of zero. Since the Gaia distance error is greater than 10\%, we do not impose an empirical stellar radius constraint. 

\subsubsection{Global Model Results\label{sec:globalresults}}

We adopt the posterior median parameter values and uncertainties of the YY-constrained fit as the fiducial global model and compare to the results from the Torres-constrained global model. The posterior median parameter values and 68\% confidence intervals for both final global models are shown in Table \ref{tab:parameters}. The KELT-19Ab fiducial model indicates the system has a host star with mass $\Mstar = 1.62 \msun$, radius $\Rstar = 1.830 \rsun$, and effective temperature $\teff = 7,500$\,K, and a planet with $\teq=1935$\,K, and radius $\RP = 1.891 \rj$. Because an RV orbit is not detected, we state $3\sigma$ upper limits on all of the planet mass related posterior parameter values. KELT-19Ab's planet mass is constrained to be $<4.07$\MJ\ with $3\sigma$ significance.

In summary, we find that the YY and Torres stellar constraints result in system parameters that are well within $1\sigma$.


\begin{table*}
\begin{center}
\caption{Global fit posterior parameter values for the KELT-19A\textrm{\normalfont b} system}
\label{tab:parameters}
\begin{tabular}{lccccc}
\hline
\hline
\multicolumn{1}{l}{~~~Parameter} & \multicolumn{1}{l}{Units} & \multicolumn{1}{c}{YY Circular (adopted)} & \multicolumn{1}{c}{Torres Circular}\\
\multicolumn{1}{l}{}             & \multicolumn{1}{l}{}      & \multicolumn{1}{c}{68\% Confidence}   & \multicolumn{1}{c}{68\% Confidence} \\
\multicolumn{1}{l}{}             & \multicolumn{1}{l}{}      & \multicolumn{1}{c}{\color{red}(99.7\% Upper Limit)}      & \multicolumn{1}{c}{\color{red}(99.7\% Upper Limit)} \\
\hline
\multicolumn{3}{l}{Stellar Parameters:}\\
                               ~~~$M_{*}$\dotfill &Mass (\msun)\dotfill & $1.62_{-0.20}^{+0.25}$             & $1.64_{-0.15}^{+0.19}$     \\
                             ~~~$R_{*}$\dotfill &Radius (\rsun)\dotfill & $1.830\pm0.099$                    & $1.832_{-0.080}^{+0.086}$     \\
                         ~~~$L_{*}$\dotfill &Luminosity (\lsun)\dotfill & $9.5_{-1.1}^{+1.2}$                & $9.5_{-1.0}^{+1.1}$     \\
                             ~~~$\rho_*$\dotfill &Density (cgs)\dotfill & $0.376_{-0.027}^{+0.031}$          & $0.378_{-0.027}^{+0.031}$     \\
                  ~~~$\log{g_*}$\dotfill &Surface gravity (cgs)\dotfill & $4.127\pm0.029$                    & $4.129\pm0.026$     \\
                  ~~~$\teff$\dotfill &Effective temperature (K)\dotfill & $7500\pm110$                       & $7500\pm110$     \\
                                 ~~~$\feh$\dotfill &Metallicity\dotfill & $-0.12\pm0.51$                     & $-0.12_{-0.34}^{+0.58}$     \\
             ~~~$v\sin{I_*}$\dotfill &Rotational velocity (m/s)\dotfill & $84800\pm2000$                     & $84800\pm2100$     \\
               ~~~$NRLW$\dotfill &Non-rotating line width (m/s)\dotfill & $3100\pm1800$                      & $3100\pm1800$     \\
\multicolumn{3}{l}{Planetary Parameters:}\\         
                                 ~~~$M_{P}$\dotfill &Mass (\mj)\dotfill & $\color{red}(<4.07)$               & $\color{red}(<4.15)$ \\
                               ~~~$R_{P}$\dotfill &Radius (\rj)\dotfill & $1.91\pm0.11$                      & $1.909_{-0.091}^{+0.06}$     \\
                           ~~~$\rho_{P}$\dotfill &Density (cgs)\dotfill & $\color{red}(<0.744)$              & $\color{red}(<0.739)$ \\
                      ~~~$\log{g_{P}}$\dotfill &Surface gravity (cgs)\dotfill & $\color{red}(<3.44)$         & $\color{red}(<3.44))$\\
               ~~~$T_{eq}$\dotfill &Equilibrium temperature (K)\dotfill & $1935\pm38$                        & $1934\pm37$     \\
                           ~~~$\Theta$\dotfill &Safronov number\dotfill & $0.0083_{-0.0071}^{+0.039}$        & $0.0083_{-0.0070}^{+0.039}$     \\
                   ~~~$\fave$\dotfill &Incident flux (\fluxcgs)\dotfill & $3.18\pm0.25$                      & $3.18\pm0.25$     \\
\multicolumn{3}{l}{Orbital Parameters:}\\
 ~~~$T_{C_0}$\dotfill &Reference time of transit from TTVs (\bjdtdb)\dotfill & $2457281.249537\pm0.000361$   & $2457281.249520\pm0.000359$     \\
      ~~~$T_{S_0}$\dotfill &Reference time of secondary eclipse (\bjdtdb)\dotfill & $2457278.94367\pm0.00036$          & $2457283.55539\pm0.00035$ \\
                        ~~~$P$\dotfill &Period from TTVs (days)\dotfill & $4.6117093\pm0.0000088$            & $4.6117093\pm0.0000089$     \\
                        ~~~$a$\dotfill &Semi-major axis (AU)   \dotfill & $0.0637_{-0.0027}^{+0.0031}$       & $0.0640_{-0.0020}^{+0.0024}$     \\
           ~~~$\lambda$\dotfill &Spin-orbit alignment (degrees)\dotfill & $-179.7_{-3.8}^{+3.7}$             & $-179.9\pm-3.8$     \\
\multicolumn{3}{l}{RV Parameters:}\\ 
                        ~~~$K$\dotfill &RV semi-amplitude (m/s)\dotfill & $\color{red}(<352)$                & $\color{red}(<355)$\\
                    ~~~$M_P\sin{i}$\dotfill &Minimum mass (\mj)\dotfill & $\color{red}(<4.05)$               & $\color{red}(<4.14)$\\
                           ~~~$M_{P}/M_{*}$\dotfill &Mass ratio\dotfill & $\color{red}(<0.00237)$            & $\color{red}(<0.00236)$\\
                       ~~~$u$\dotfill &RM linear limb darkening\dotfill & $0.5440_{-0.0059}^{+0.014}$        & $0.5460_{-0.0076}^{+0.017}$     \\
                            ~~~$\gamma_{McDonald}$\dotfill &m/s\dotfill & $-7256\pm90$                       & $-7258\pm90$    \\
                                ~~~$\gamma_{TRES}$\dotfill &m/s\dotfill & $-8150\pm180$                      & $-8150\pm180$    \\
\multicolumn{3}{l}{Primary Transit Parameters:}\\
~~~$R_{P}/R_{*}$\dotfill &Radius of the planet in stellar radii\dotfill & $0.10713\pm0.00092$                & $0.10709\pm0.00093$    \\
           ~~~$a/R_*$\dotfill &Semi-major axis in stellar radii\dotfill & $7.50_{-0.18}^{+0.20}$             & $7.52\pm0.20$    \\
                          ~~~$i$\dotfill &Inclination (degrees)\dotfill & $85.41_{-0.31}^{+0.34}$            & $85.34_{-0.32}^{+0.35}$    \\
                               ~~~$b$\dotfill &Impact parameter\dotfill & $0.601_{-0.030}^{+0.026}$          & $0.599_{-0.031}^{+0.026}$    \\
                             ~~~$\delta$\dotfill &Transit depth\dotfill & $0.01148\pm0.00020$                & $0.01147\pm0.00020$    \\
                    ~~~$T_{FWHM}$\dotfill &FWHM duration (days)\dotfill & $0.15645\pm0.00075$                & $0.15650\pm0.00076$    \\
              ~~~$\tau$\dotfill &Ingress/egress duration (days)\dotfill & $0.0266\pm0.0016$                  & $0.0265\pm0.0016$    \\
                     ~~~$T_{14}$\dotfill &Total duration (days)\dotfill & $0.1831\pm0.0015$                  & $0.1830\pm0.0015$    \\
   ~~~$P_{T}$\dotfill &A priori non-grazing transit probability\dotfill & $0.1190\pm0.0030$                  & $0.1188\pm0.0030$    \\
             ~~~$P_{T,G}$\dotfill &A priori transit probability\dotfill & $0.1476_{-0.0039}^{+0.0037}$       & $0.1473_{-0.0039}^{+0.0038}$     \\
                     ~~~$u_{1B}$\dotfill &Linear Limb-darkening\dotfill & $0.3798_{-0.0092}^{+0.018}$        & $0.382_{-0.011}^{+0.022}$\\
                  ~~~$u_{2B}$\dotfill &Quadratic Limb-darkening\dotfill & $0.3483_{-0.011}^{+0.0071}$        & $0.3487_{-0.012}^{+0.0064}$\\
                     ~~~$u_{1I}$\dotfill &Linear Limb-darkening\dotfill & $0.139_{-0.011}^{+0.034}$          & $0.139_{-0.011}^{+0.040}$     \\
                  ~~~$u_{2I}$\dotfill &Quadratic Limb-darkening\dotfill & $0.319_{-0.031}^{+0.018}$          & $0.319_{-0.034}^{+0.021}$     \\
                ~~~$u_{1Sloang}$\dotfill &Linear Limb-darkening\dotfill & $0.3500_{-0.0082}^{+0.022}$        & $0.3513_{-0.0089}^{+0.026}$    \\
             ~~~$u_{2Sloang}$\dotfill &Quadratic Limb-darkening\dotfill & $0.344_{-0.015}^{+0.012}$          & $0.344_{-0.016}^{+0.013}$     \\
                ~~~$u_{1Sloani}$\dotfill &Linear Limb-darkening\dotfill & $0.1558_{-0.0100}^{+0.037}$        & $0.156_{-0.010}^{+0.043}$     \\
             ~~~$u_{2Sloani}$\dotfill &Quadratic Limb-darkening\dotfill & $0.324_{-0.032}^{+0.018}$          & $0.324_{-0.036}^{+0.021}$     \\
                ~~~$u_{1Sloanr}$\dotfill &Linear Limb-darkening\dotfill & $0.2221_{-0.0066}^{+0.036}$        & $0.2221_{-0.0061}^{+0.042}$\\
             ~~~$u_{2Sloanr}$\dotfill &Quadratic Limb-darkening\dotfill & $0.347_{-0.030}^{+0.014}$          & $0.347_{-0.034}^{+0.016}$\\
                ~~~$u_{1Sloanz}$\dotfill &Linear Limb-darkening\dotfill & $0.109_{-0.013}^{+0.026}$          & $0.109_{-0.013}^{+0.030}$     \\
             ~~~$u_{2Sloanz}$\dotfill &Quadratic Limb-darkening\dotfill & $0.311_{-0.025}^{+0.018}$          & $0.311_{-0.027}^{+0.021}$     \\
\hline
\end{tabular}
\end{center}
\end{table*}

\subsubsection{Transit Timing Variation Results}\label{sec:ttvs}

\begin{figure}
\centering \includegraphics[width=\columnwidth, trim=0.0cm 0.0cm 0.0cm 0.0cm, clip=true]{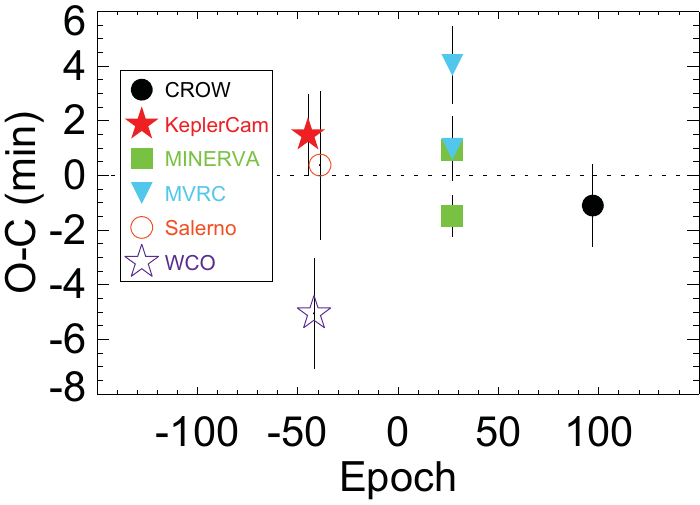}
\caption{KELT-19Ab transit timing variations. All of the timing deviations are less than $3\sigma$ from the linear ephemeris. The transit center times of the four transits on epoch 27 range from $\sim-2\sigma$ to $+3\sigma$, indicating that the TTVs are likely due to light curve systematics rather than astrophysical influences.}
\label{fig:ttv}
\end{figure}

 We derive a precise linear ephemeris from the transit timing data by fitting a straight line to all inferred transit center times.  These times are listed in Table \ref{tbl:ttv} and plotted in Figure \ref{fig:ttv}. We find a best fit linear ephemeris of $T_{\rm 0} = 2457281.249537 \pm 0.000362$ \bjdtdb, $P_{\rm Transits} = 4.6117091 \pm 9.0\times10^{-6}$\,d, with a $\chi^2$ of 20.9 and 6 degrees of freedom, resulting in $\chisqr=3.5$. While the $\chisqr$ is larger than one might expect, this is often the case in ground-based TTV studies, likely due to systematics in the transit data. Even so, all of the timing deviations are less than $3\sigma$ from the linear ephemeris. Furthermore, note that the TTVs of the four simultaneous transit observations on epoch 27 range from $\sim-2\sigma$ to $+3\sigma$, indicating that the TTVs are likely due to light curve systematics.  We therefore conclude that there is no convincing evidence for TTVs in the KELT-19Ab system. However, due to the limited number of full light curves included in this study, we suggest further transit observations of KELT-19Ab before ruling out TTVs.

\begin{table}
 \centering
 \caption{Transit times for KELT-19A\textrm{\normalfont b}.}
 \label{tbl:ttv}
 \begin{tabular}{r@{\hspace{12pt}} l r r r c}
    \hline
    \hline
    \multicolumn{1}{c}{Epoch} & \multicolumn{1}{c}{$T_{C}$} 	& \multicolumn{1}{l}{$\sigma_{T_C}$} 	& \multicolumn{1}{c}{O-C} &  \multicolumn{1}{c}{O-C} 			& Telescope \\
	    & \multicolumn{1}{c}{(\bjdtdb)} 	& \multicolumn{1}{c}{(s)}			& \multicolumn{1}{c}{(s)} &  \multicolumn{1}{c}{($\sigma_{T_{C}}$)} 	& \\
    \hline
 -45 & 2457073.723660 &  90 &  89.10 &  0.98 &  KeplerCam  \\
 -42 & 2457087.554255 & 122 &-302.48 & -2.48 &  WCO  \\
 -39 & 2457101.393149 & 163 &  22.97 &  0.14 &  Salerno  \\
  27 & 2457405.764653 &  45 & -88.84 & -1.97 &  MINERVA  \\
  27 & 2457405.766335 &  59 &  56.49 &  0.96 &  MINERVA  \\
  27 & 2457405.768490 &  86 & 242.68 &  2.80 &  MVRC  \\
  27 & 2457405.766362 &  71 &  58.82 &  0.82 &  MVRC  \\
  97 & 2457728.584553 &  90 & -65.88 & -0.73 &  CROW  \\
    \hline
    \hline
 \end{tabular}
\end{table}

\section{False Positive Analysis}

Despite the lack of a definitive measurement of the companion mass, we are confident that this system is truly a hot Jupiter transiting a late A star.  The evidence for this comes from several sources which we will briefly review. However, we invite the reader to review papers by \citet{Bieryla:2015}, \citet{Zhou:2016,Zhou:2017}, and \citet{Hartman:2015} for a more detailed explanation.  The basic point is that the radial velocity measurements, while not sufficiently precise to measure the mass of the transiting companion, do indicate that it is not a brown dwarf or a low-mass star, if it is indeed transiting the primary A star.  On the other hand, we are confident that the companion is transiting the primary A star (rather than, say, the later spectral-type bound companion), because we see a Doppler tomographic signal with the expected amplitude, duration, and impact parameter inferred from the follow-up light curves.  Of course, the first system to have been validated in this way was WASP-33b \citep{CollierCameron2010}.

The Doppler tomographic observation eliminates the possibility of a blended eclipsing binary causing the transit signal. Even though the line profile derived from the least-squares deconvolution shows a spectroscopic companion blended with KELT-19A, the spectroscopic transit is seen crossing nearly the entirety of the rapidly rotating primary star's line profile (the TRES DT observations did not cover ingress), confirming that the planet is indeed orbiting KELT-19A. The summed flux underneath the Doppler tomographic shadow and the distance of closest approach of the shadow from the zero velocity at the center of the predicted transit time is consistent with both the photometric transit depth and impact parameter, suggesting that the photometric transit is not diluted by background stars, and is fully consistent with the spectroscopic transit. 

Adaptive optics observations (\S \ref{sec:ao}) also show a nearby companion consistent in relative brightness with the TRES companion's relative brightness, but no other stars brighter than $\Delta Br$-$\gamma<6$ with separation $>0.6\arcsec$ from KELT-19A at $5\sigma$ significance. Furthermore, the deblended follow-up observation transit depths are consistent across the optical and infrared bands as indicated in Figure \ref{fig:primary_lcs}.

Finally, the planetary nature of KELT-19Ab is confirmed by the TRES and HJST radial velocity measurements, which constrain the mass of the companion to be $\la 4.1 $\,\MJ\ at $3\sigma$ significance. This eliminates the possibility that the transiting companion is a stellar or brown-dwarf-mass object. 

Thus we conclude that all the available evidence suggests that the most plausible interpretation is that KELT-19Ab is a Jupiter-size planet transiting the late A-star TYC 764-1494-1 with a retrograde projected spin-orbit alignment (see \S \ref{sec:dopptom} and \S \ref{sec:spin-orbit-misalignment}), and with a late G or early K bound companion with a projected separation of $\approx 160$\,AU.

We do note, however, that this was a particularly complicated case; one that may have easily been rejected as a false positive based simply on the double-lined nature of the line profiles (see Figure \ref{fig:lsdfit}).  KELT-19Ab therefore provides an important object lesson: transiting planets can indeed be found and definitively confirmed in initially unresolved binary systems.  Indeed, such systems may provide important constraints on the emplacement of hot Jupiters, as the outer bound stellar companion can easily be responsible for Kozai-Lidov oscillations and so emplacement of hot Jupiters \citep{Kozai:1962,Lidov:1962}.  

\section{Discussion}

\begin{figure}
\vspace{0.1in}
\includegraphics[width=1\linewidth]{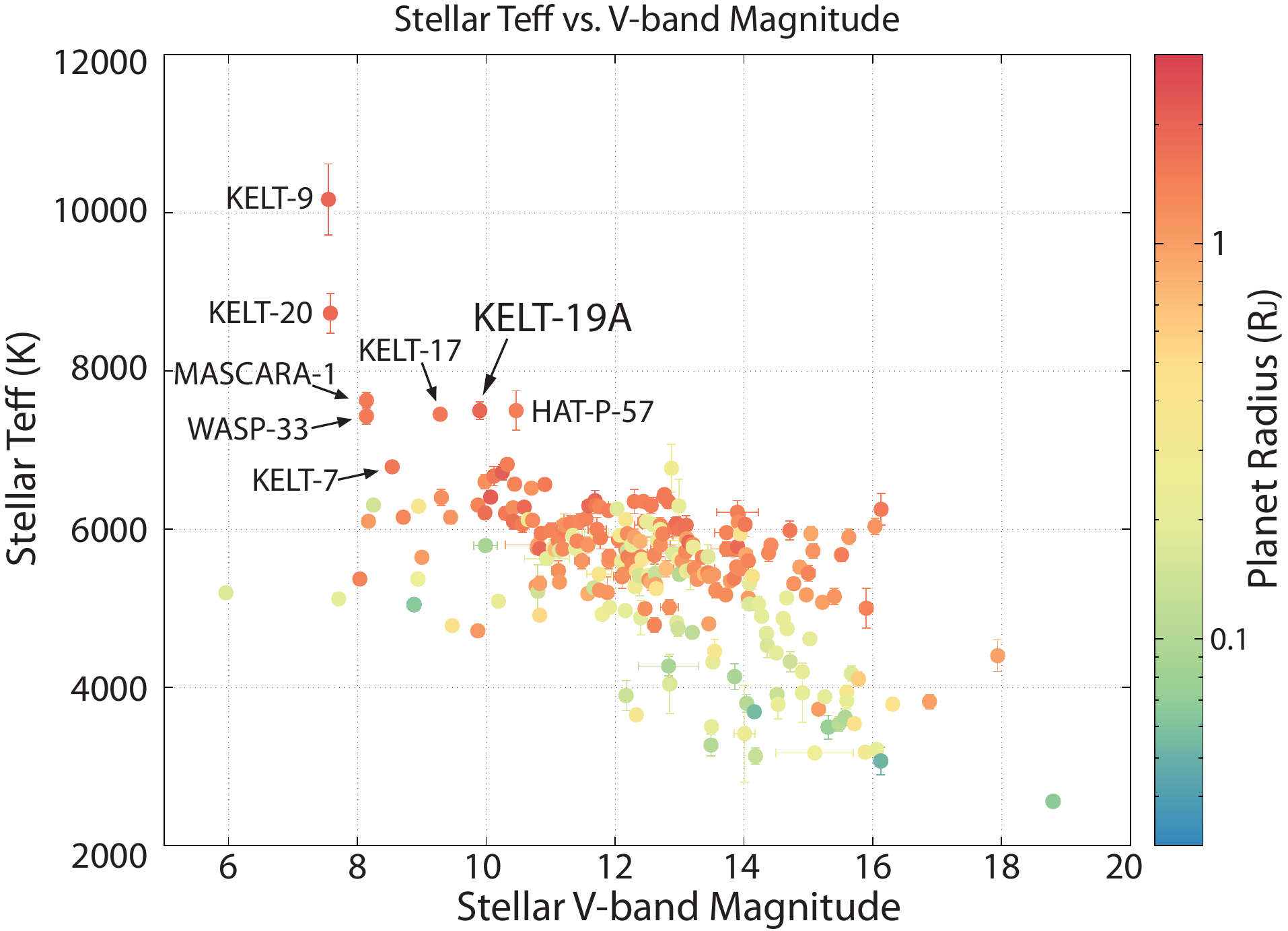}
\caption{The population of transiting exoplanets based on the host star's V-band magnitude and effective temperature (\teff), with colors indicating the radius of the planet in \rj. Within \teff uncertainties, KELT-19A joins KELT-17, HAT-P-57, and WASP-33 as having the third highest \teff of all known transiting hot Jupiter host stars. These data, except KELT-19A and KELT-20, were extracted from the NASA Exoplanet Database\footnote{https://exoplanetarchive.ipac.caltech.edu}.}
\label{fig:Teff_Mag}
\end{figure}

Figure \ref{fig:Teff_Mag} shows host star effective temperature versus V-band magnitude for known transiting planets. Within \teff uncertainties, KELT-19A joins KELT-17, WASP-33, HAT-P-57, and MASCARA-1 as having the third highest \teff of all known transiting hot Jupiter host stars. With a host star luminosity of $\sim 9.5\lsun$ and an orbital period of $\sim 4.6$~days, the planet has a high equilibrium temperature of $\teq\sim 2000$\,K, assuming zero albedo and perfect heat redistribution.  With a V-band magnitude of $9.9$, a high equilibrium temperature, and a likely large scale height, it is an excellent target for detailed follow-up and characterization. Because KELT-19A is an A star, the planet receives a higher amount of high-energy radiation than the majority of known transiting planet systems, which may lead to significant atmospheric ablation \citep{Murray-Clay:2009}.

\subsection{Spin-Orbit Misalignment}\label{sec:spin-orbit-misalignment}

Although we have measured the sky-projected spin-orbit misalignment $\lambda$, we cannot measure the full three-dimensional spin-orbit misalignment $\psi$ because we do not know the inclination angle of the stellar rotation axis with respect to the line of sight, $I_*$. We can, however, set limits upon $\psi$. First, we follow \cite{Iorio:2011} and limit $I_*$, and therefore $\psi$, by requiring that KELT-19A must rotate at less than break-up velocity. Doing so, we find that, at $1\sigma$ confidence, $19.7^{\circ}<I_*<160.0^{\circ}$ and $105^{\circ}<\psi<180^{\circ}$. We can, however, use the possible Am star nature of KELT-19A to set somewhat stricter limits upon $I_*$ and $\psi$. Although physically KELT-19A must have an equatorial rotation velocity of $v_{\mathrm{eq}}<250$\kms\ to avoid break-up, empirically Am stars are not observed to have rotation velocities of greater than $\sim150$\kms. If we instead require that KELT-19A have $v_{\mathrm{eq}}<150$\kms, we obtain limits of $33.5^{\circ}<I_*<146.5^{\circ}$ and $119^{\circ}<\psi<180^{\circ}$. 

KELT-19Ab continues the trend of hot Jupiters around A stars to have a wide range of sky-projected spin-orbit misalignments. Of the eight A-star-hosted hot Jupiters with measured spin-orbit misalignments to date, one is on a prograde, well-aligned orbit \citep[KELT-20b/MASCARA-2b:][]{Lund:2017,Talens:2017c}; two have misaligned prograde orbits \citep[Kepler-13Ab and MASCARA-1b:][]{Johnson:2014,Talens:2017b}; one is in a prograde orbit with an unclear degree of misalignment \citep[HAT-P-57b:][]{Hartman:2015}; one is on a near-polar orbit \citep[KELT-9b:][]{Gaudi:2017}; two are on misaligned retrograde orbits \citep[WASP-33b and KELT-17b:][]{CollierCameron2010,Zhou:2016}; and, now, KELT-19Ab is on a near-antialigned retrograde orbit. Qualitatively, the distribution of A-star hot Jupiter spin-orbit misalignments appears consistent with isotropic, but detailed investigation of this distribution will require a larger sample of planets.

\subsection{Tidal Evolution and Irradiation History}\label{sec:Irradiation}

Following \citet{Penev:2014}, we model the orbital evolution of KELT-19Ab due to the dissipation of the tides raised by the planet on the the host star under the assumption of a constant phase lag. The starting configuration of the system was tuned to reproduce the presently observed system parameters (Table \ref{tab:parameters}) at the assumed system age of 1.1 Gyr (see \S \ref{sec:hrd_and_age}). The evolution model includes the effects of the changing stellar radius and luminosity following the YY circular stellar model with mass and metallicity as given in Table \ref{tab:parameters}. No effects of the stellar rotation have been included in the calculation, since the star is observed to counter-rotate with respect to the orbit. In a retrograde configuration, tidal coupling always acts to remove energy and angular momentum from the planet, and as a result under the assumption of a constant phase lag, the evolution is indistinguishable from that of a non-rotating host star.

\begin{figure}
\includegraphics[width=1\linewidth]{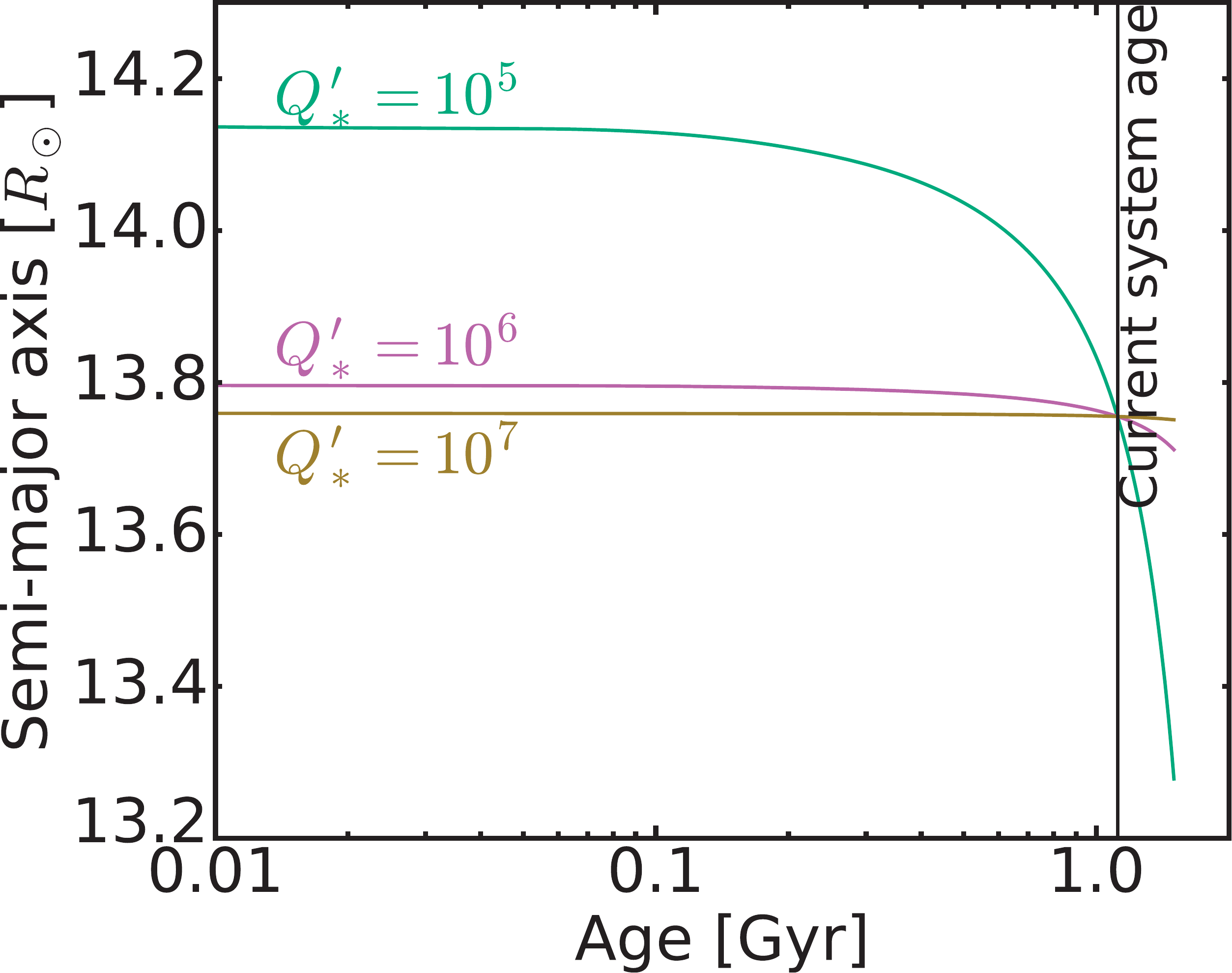}
\includegraphics[width=1\linewidth]{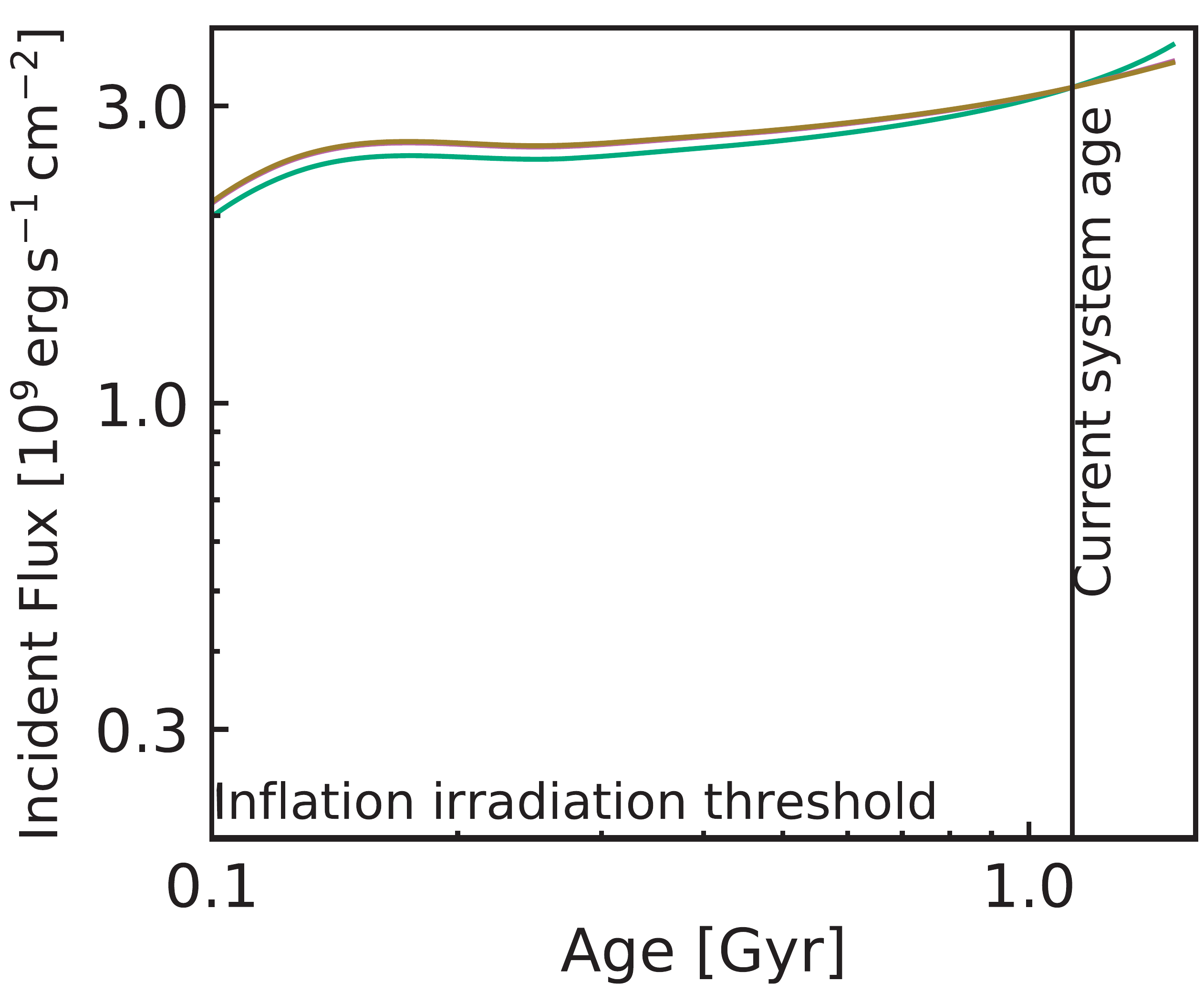}
   \vspace{-.2in}
   \caption{\footnotesize
   {\tt (Top)} The orbital semi-major axis history of KELT-19Ab modeled for a range of stellar tidal quality factors, $Q_*'$, where $Q_*'^{-1}$ is the product of the tidal phase lag and the Love number. The black vertical line marks the current system age of 1.1 Gyr. {\tt (Bottom)} The irradiation history of KELT-19Ab modeled for a range of stellar tidal quality factors. The black horizontal line nearly coincident with the $x$-axis marks the inflation irradiation threshold of $\approx 2 \times 10^8$ erg s$^{-1}$ cm$^{-2}$  \citep{Demory:2011}}.
   \label{fig:Tides}
\end{figure}

Orbital and stellar irradiation evolutions are shown in Figure \ref{fig:Tides} for a range of stellar tidal quality factors ($Q_*' = 10^5, 10^6, \textrm{and}\, 10^7$), where $Q_*'^{-1}$ is the product of the tidal phase lag and the Love number. We find that the insolation received by the planet is well above the empirical inflation irradiation threshold of $\sim 2 \times 10^8$ erg s$^{-1}$ cm$^{-2}$ \citep{Demory:2011} for the entire main-sequence existence of the star (bottom panel of Figure \ref{fig:Tides}).

We consider a wide range of $Q_*'$ because of the wide range of proposed mechanisms for tidal dissipation in current theoretical models and the conflicting observational constraints backing those models, especially for stars that may have surface convective zones (see the review by \citealt{Ogilvie:2014} and references therein). Furthermore, because the dependence on stellar mass and tidal frequency is different for the different proposed mechanisms, we make the simplifying assumption that $Q_*'$ remains constant over the life of the star. However, with multi-year baselines, it may be possible in the future to empirically constrain the lower limit on $Q_*'$ for KELT-19Ab via precise measurements of the orbital period time decay (cf. \citealt{Hoyer:2016}).

Finally, note that this model does not account in any way for the larger-distance Type II or scattering-induced migration that KELT-19Ab and other hot Jupiters likely undergo. It considers only the close-in migration due to tidal friction alone.

\section{Conclusion}

KELT-19 consists of a hierarchical triple system of an Am star that is being transited by a $P\sim 4.6$~day hot Jupiter with a mass of $\la 4~\mj$. The planet is highly inflated and highly irradiated, with a radius of $\simeq 2~\rj$, and an equilibrium temperature of $\teq\sim 2000$K.  It is also on a retrograde orbit with projected spin-orbit alignment of $\lambda \sim -180$~degrees.  Finally, the primary A star (KELT-19A) and hot Jupiter (KELT-19Ab) are orbited by an outer bound stellar G9V/K1V companion (KELT-19B) with a projected separation of $\sim 160$~AU.  

In many ways, KELT-19 is one of the most unusual transiting hot Jupiter systems yet discovered.  Firstly, the primary star (KELT-19A) and planet host is an Am (metallic line-enhanced) star. To the best of our knowledge, this is the only such star known to host a transiting hot Jupiter\footnote{However, see \citealt{Genier:1999}, who suggest that WASP-33 may be an Am star, although \citealt{CollierCameron2010} note that ``No obvious Am characteristics are visible in this spectrum other than slightly weak Ca II H\&K lines''}. As is the case for other Am stars, KELT-19A rotates slowly compared to stars of similar effective temperature. Although the presence of a nearby stellar companion is usually invoked to explain both the slower rotation and peculiar abundance patterns of Am stars, the stellar companion KELT-19B seems too distant to cause significant tidal braking.  Furthermore, the planetary companion (KELT-19Ab) is likely too low mass to sufficiently slow the rotation of its host star, KELT-19A \citep{Matsumura:2010}. Thus, we believe that the slow rotation of KELT-19A is either primordial or was induced by a more efficient tidal braking mechanism than expected.

Finally, we note that the confirmation of KELT-19Ab provides an important object lesson for future transit surveys.  The initial line-spread function exhibited two peaks: a broad peak due to the rapidly rotating A star, and a narrower peak due to the more slow-rotating (but bound) blended late G/early K companion. Without careful analysis, such multiple-star systems may be spuriously rejected as false positives. Generally, we suggest that multi-lined systems not be immediately discarded unless the line of the blended secondary shows relative motion that is consistent with the photometric ephemeris of the transit event, in which case the secondary is likely one component of a eclipsing binary, whose eclipses are being diluted by the primary. In this case, our analysis revealed the presence of a genuine transiting hot Jupiter orbiting an A-type star in a hierarchical triple system.

\acknowledgements
This project makes use of data from the KELT survey, including support from The Ohio State University, Vanderbilt University, and Lehigh University, along with the KELT follow-up collaboration.
Work performed by J.E.R. was supported by the Harvard Future Faculty Leaders Postdoctoral fellowship.
D.J.S. and B.S.G. were partially supported by NSF CAREER Grant AST-1056524.
Work by S.V.Jr. is supported by the National Science Foundation Graduate Research Fellowship under Grant No. DGE-1343012.
Work by G.Z. is provided by NASA through Hubble Fellowship grant HST-HF2-51402.001-A awarded by the Space Telescope Science Institute, which is operated by the Association of Universities for Research in Astronomy, Inc., for NASA, under contract NAS 5-26555.
This paper includes data taken at The McDonald Observatory of The University of Texas at Austin.
This work has made use of NASA's Astrophysics Data System, the Extrasolar Planet Encyclopedia, the NASA Exoplanet Archive, the SIMBAD database operated at CDS, Strasbourg, France, and the VizieR catalogue access tool, CDS, Strasbourg, France.  We make use of Filtergraph, an online data visualization tool developed at Vanderbilt University through the Vanderbilt Initiative in Data-intensive Astrophysics (VIDA).
We also used data products from the Widefield Infrared Survey Explorer, which is a joint project of the University of California, Los Angeles; the Jet Propulsion Laboratory/California Institute of Technology, which is funded by the National Aeronautics and Space Administration; the Two Micron All Sky Survey, which is a joint project of the University of Massachusetts and the Infrared Processing and Analysis Center/California Institute of Technology, funded by the National Aeronautics and Space Administration and the National Science Foundation; and the European Space Agency (ESA) mission {\it Gaia} (\url{http://www.cosmos.esa.int/gaia}), processed by the {\it Gaia} Data Processing and Analysis Consortium (DPAC, \url{http://www.cosmos.esa.int/web/gaia/dpac/consortium}). Funding for the DPAC has been provided by national institutions, in particular the institutions participating in the {\it Gaia} Multilateral Agreement. MINERVA is a collaboration among the Harvard-Smithsonian Center for Astrophysics, The Pennsylvania State University, the University of Montana, and the University of New South Wales. MINERVA is made possible by generous contributions from its collaborating institutions and Mt. Cuba Astronomical Foundation, The David \& Lucile Packard Foundation, National Aeronautics and Space Administration (EPSCOR grant NNX13AM97A), The Australian Research Council (LIEF grant LE140100050), and the National Science Foundation (grants 1516242 and 1608203). Any opinions, findings, and conclusions or recommendations expressed are those of the author and do not necessarily reflect the views of the National Science Foundation. This work was partially supported by funding from the Center for Exoplanets and Habitable Worlds. The Center for Exoplanets and Habitable Worlds is supported by the Pennsylvania State University, the Eberly College of Science, and the Pennsylvania Space Grant Consortium.

\bibliographystyle{apj}

\bibliography{KELT-19b}

\end{document}